\documentclass[12pt, preprint, longabstract,usegraphicx]{aastex}
\usepackage{lscape}

\newcommand\msun {M_\odot}
\newcommand\mearth {{M_\oplus}}
\newcommand\gtsima{$\; \buildrel >\over\sim \;$}
\newcommand\simgt{\lower.5ex\hbox{\gtsima}}

\begin{document}
\title{A Cold Neptune-Mass Planet OGLE-2007-BLG-368Lb: Cold Neptunes Are Common}

\author{
T.~Sumi\altaffilmark{1,2},
D.P.~Bennett\altaffilmark{1,3,4},
I.A.~Bond\altaffilmark{1,5},
A.~Udalski\altaffilmark{6,7},
V.~Batista\altaffilmark{3,8,9},
M.~Dominik\altaffilmark{3,10,11},
P.~Fouqu\'e\altaffilmark{3,12},
D.~Kubas\altaffilmark{3,8,9,13},
A.~Gould\altaffilmark{14,15},
B.~Macintosh\altaffilmark{16},
K.~Cook\altaffilmark{16},
S.~Dong\altaffilmark{14,17},
L.~Skuljan\altaffilmark{1,5},
A.~Cassan\altaffilmark{3,8,9,18},
 \\ and \\
F.~Abe\altaffilmark{2},
C.S.~Botzler\altaffilmark{19},
A.~Fukui\altaffilmark{2},
K.~Furusawa\altaffilmark{2},
J.B.~Hearnshaw\altaffilmark{20},
Y.~Itow\altaffilmark{2},
K.~Kamiya\altaffilmark{2},
P.M.~Kilmartin\altaffilmark{21},
A.~Korpela\altaffilmark{22},
W.~Lin\altaffilmark{5},
C.H.~Ling\altaffilmark{5},
K.~Masuda\altaffilmark{2},
Y.~Matsubara\altaffilmark{2},
N.~Miyake\altaffilmark{2},
Y.~Muraki\altaffilmark{23},
M.~Nagaya\altaffilmark{2},
T.~Nagayama\altaffilmark{24},
K.~Ohnishi\altaffilmark{25},
T.~Okumura\altaffilmark{2},
Y.C.~Perrott\altaffilmark{19}
N.~Rattenbury\altaffilmark{19},
To.~Saito\altaffilmark{26},
T.~Sako\altaffilmark{2},
D.J.~Sullivan\altaffilmark{22},
W.L.~Sweatman\altaffilmark{5},
P.,J.~Tristram\altaffilmark{21},
P.C.M.~Yock\altaffilmark{19}\\
({The MOA Collaboration}),\\
J.P.~Beaulieu\altaffilmark{8,9}
A.~Cole\altaffilmark{27},
Ch.~Coutures\altaffilmark{8,9}
M.F.~Duran\altaffilmark{28},
J.~Greenhill\altaffilmark{27},
F.~Jablonski\altaffilmark{29},
U.~Marboeuf\altaffilmark{30}, 
E.~Martioli\altaffilmark{29}, 
E.~Pedretti\altaffilmark{10},  
O.~Pejcha\altaffilmark{15},  
P.~Rojo\altaffilmark{28},
M.D.~Albrow\altaffilmark{20}, 
S.~Brillant\altaffilmark{14}, 
M.~Bode\altaffilmark{31},
D.M.~Bramich\altaffilmark{32},
M.J.~Burgdorf\altaffilmark{33,34},  
J.A.R.~Caldwell\altaffilmark{35}, 
H.~Calitz\altaffilmark{36},
E.~Corrales\altaffilmark{8,9}, 
S.~Dieters\altaffilmark{8,9,27},  
D.~Dominis Prester\altaffilmark{37},
J.~Donatowicz\altaffilmark{8,38},  
K.~Hill\altaffilmark{9,27},
M.~Hoffman\altaffilmark{36}, 
K.~Horne\altaffilmark{10},  
U.G.~J{\o}rgensen\altaffilmark{39},
N.~Kains\altaffilmark{10},
S.~Kane\altaffilmark{40}, 
J.B.~Marquette\altaffilmark{8,9}, 
R.~Martin\altaffilmark{41},
P.~Meintjes\altaffilmark{36},  
J.~Menzies\altaffilmark{42}, 
K.R.~Pollard\altaffilmark{20},
K.C.~Sahu\altaffilmark{43}, 
C.~Snodgrass\altaffilmark{14}, 
I.~Steele\altaffilmark{31}, 
R.~Street\altaffilmark{44}, 
Y.~Tsapras\altaffilmark{44}, 
J.~Wambsganss\altaffilmark{18},
A.~Williams\altaffilmark{41}, 
M.~Zub\altaffilmark{18}\\
({The PLANET Collaboration}),\\
M.K.~Szyma{\'n}ski\altaffilmark{7},
M.~Kubiak\altaffilmark{7},
G.~Pietrzy{\'n}ski\altaffilmark{7,45},
I.~Soszy{\'n}ski\altaffilmark{7},
O.~Szewczyk\altaffilmark{45},
{\L}.~Wyrzykowski\altaffilmark{46},
{K}.~Ulaczyk\altaffilmark{7}\\
(The OGLE Collaboration),\\
W.~Allen\altaffilmark{47},
G.W.~Christie\altaffilmark{48},
D.L.~DePoy\altaffilmark{49},
B.S.~Gaudi\altaffilmark{15},
C.~Han\altaffilmark{50},
J.~Janczak\altaffilmark{15},
C.-U.~Lee\altaffilmark{51},
J.~McCormick\altaffilmark{52},
F.~Mallia\altaffilmark{53}
B.~Monard\altaffilmark{54},
T.~Natusch\altaffilmark{55},
B.-G.~Park\altaffilmark{51},
R.W.~Pogge\altaffilmark{15}\\
R.~Santallo\altaffilmark{56}\\
(The $\mu$FUN Collaboration),\\
}
\altaffiltext{1}
{Microlensing Observations in Astrophysics (MOA)}
\altaffiltext{2}{Solar-Terrestrial Environment Laboratory, Nagoya University, Nagoya, 464-8601, Japan; sumi,abe,afukui,furusawa,itow,kkamiya,kmasuda,ymatsu,nmiyake,mnagaya,okumurat,sako@stelab.nagoya-u.ac.jp}
\altaffiltext{3}
{Probing Lensing Anomalies NETwork (PLANET)}
\altaffiltext{4}
{Department of Physics, University of Notre Dame, Notre Dame, IN 46556, USA; bennett@nd.edu}
\altaffiltext{5}
{Institute of Information and Mathematical Sciences, Massey University,
Private Bag 102-904, North Shore Mail Centre, Auckland, New Zealand;
i.a.bond,l.skuljan,w.lin,c.h.ling,w.sweatman@massey.ac.nz}
\altaffiltext{6}
{Optical Gravitational Lens Experiment (OGLE)}
\altaffiltext{7}
{Warsaw University Observatory, Al.~Ujazdowskie~4, 00-478~Warszawa,Poland; udalski, msz, mk, pietrzyn, soszynsk, szewczyk, kulaczyk@astrouw.edu.pl}
\altaffiltext{8} 
{HOLMES Collaboration}
\altaffiltext{9}
{Institut d'Astrophysique de Paris, UMR7095 CNRS, Universit{\'e} Pierre \& Marie Curie, 98 bis boulevard Arago, 75014 Paris, France. batista,dkubas,cassan,beaulieu,coutures,marquett@iap.fr}
\altaffiltext{10}
{SUPA, Physics \& Astronomy, North Haugh, St Andrews, KY16~9SS, UK; md35,kdh1@st-andrews.ac.uk}
\altaffiltext{11} {Royal Society University Research Fellow}
\altaffiltext{12}
{Laboratoire d'Astrophysique (CNRS), Univ.\ Paul Sabatier - Toulouse 3, 14, avenue Edouard Belin, F-31400 Toulouse, France. pfouque@ast.obs-mip.fr}
\altaffiltext{13}
{European Southern Observatory, Casilla 19001, Vitacura 19, Santiago, Chile}
\altaffiltext{14}
{Microlensing Follow Up Network ($\mu$FUN)}
\altaffiltext{15}
{Department of Astronomy, Ohio State University, 140 W.\ 18th Ave., Columbus, OH 43210, USA; dong,gaudi,gould,pogge@astronomy.ohio-state.edu}
\altaffiltext{16}
{Lawrence Livermore National Laboratory, IGPP, P.O. Box 808, Livermore, CA 94551, USA}
\altaffiltext{17}
{Institute for Advanced Study, Princeton, NJ 08540, USA,\\ e-mail: {\tt dong@ias.edu}}
\altaffiltext{18}
{Astronomisches Rechen-Institut, Zentrum f\"ur~Astronomie, Heidelberg University, M\"{o}nchhofstr.~12--14, 69120 Heidelberg, Germany}
\altaffiltext{19}
{Department of Physics, University of Auckland, Private Bag 92019, Auckland, New Zealand; c.botzler,p.yock@auckland.ac.nz,yper006@aucklanduni.ac.nz}
\altaffiltext{20}
{University of Canterbury, Department of Physics and Astronomy, Private Bag 4800, Christchurch 8020, New Zealand.}
\altaffiltext{21}
{Mt.\ John Observatory, P.O. Box 56, Lake Tekapo 8770, New Zealand.}
\altaffiltext{22}
{School of Chemical and Physical Sciences, Victoria University, Wellington, New Zealand;a.korpela@niwa.co.nz,denis.sullivan@vuw.ac.nz}
\altaffiltext{23}
{Department of Physics, Konan University, Nishiokamoto 8-9-1, Kobe 658-8501, Japan.}
\altaffiltext{24}
{Department of Physics and Astrophysics, Faculty of Science, Nagoya University, Nagoya 464-8602, Japan}
\altaffiltext{25}
{Nagano National College of Technology, Nagano 381-8550, Japan.}
\altaffiltext{26}
{Tokyo Metropolitan College of Industrial Technology, Tokyo 116-8523, Japan.}
\altaffiltext{27}
{University of Tasmania, School of Maths and Physics, Private bag 37, GPO Hobart, Tasmania 7001, Australia}
\altaffiltext{28}
{Department of Astronomy, Universidad de Chile, Santiago, Chile}
\altaffiltext{29}
{Instituto Nacional de Pesquisas Espaciais, Sao Jose dos Campos, SP - Brazil}
\altaffiltext{30}
{Observatoire de Besanon, BP 1615, 25010 Besanon Cedex, France}
\altaffiltext{31} 
{Astrophysics Research Institute, Liverpool John Moores Univ., Twelve Quays House, Egerton Wharf, Birkenhead CH41 1LD, UK}
\altaffiltext{32}
{European Southern Observatory, Karl-Schwarzschild-Stra\ss{}e 2, 85748 Garching bei M\"unchen, Germany}
\altaffiltext{33}
{Deutsches SOFIA Institut, Universitaet Stuttgart, Pfaffenwaldring 31, 70569 Stuttgart, Germany}
\altaffiltext{34}
{OFIA Science Center, NASA Ames Research Center, Mail Stop N211-3, Moffett Field CA 94035, USA,mburgdorf@sofia.usra.edu}
\altaffiltext{35}
{McDonald Observatory, 16120 St Hwy Spur 78 \#2, Fort Davis, TX 79734, USA. caldwell@astro.as.utexas.edu}
\altaffiltext{36}
{Dept.\ of Physics/Boyden Observatory, University of the Free State, Bloemfontein 9300, South Africa. calitzjj.sci@mail.uovs.ac.za}
\altaffiltext{37}
{Physics department, Faculty of Arts and Sciences, University of Rijeka, 51000 Rijeka, Croatia}
\altaffiltext{38}
{Technical University of Vienna, Dept. of Computing, Wiedner Hauptstrasse 10, Vienna, Austria}
\altaffiltext{39}
{Niels Bohr Institute, Astronomical Observatory, Juliane Maries Vej 30, DK-2100 Copenhagen, Denmark}
\altaffiltext{40}
{NASA Exoplanet Science Institute, Caltech, MS 100-22, 770 South Wilson Avenue, Pasadena, CA 91125, USA}
\altaffiltext{41}
{Perth Observatory, Walnut Road, Bickley, Perth 6076, Australia. rmartin,andrew@physics.uwa.edu.au}
\altaffiltext{42}
{South African Astronomical Observatory, P.O. Box 9 Observatory 7935, South Africa}
\altaffiltext{43}
{Space Telescope Science Institute, 3700 San Martin Drive, Baltimore, MD 21218, USA, ksahu@stsci.edu}
\altaffiltext{44}
{Las Cumbres Observatory Global Telescope Network, 6740B Cortona Dr, Suite 102, Goleta, CA, 93117, USA. rstreet,ytsapras@lcogt.net}
\altaffiltext{45}
{Universidad de Concepci{\'o}n, Departamento de Fisica, Casilla 160--C, Concepci{\'o}n, Chile}
\altaffiltext{46} 
{Institute of Astronomy, Cambridge University, Madingley Rd., CB3 0HA Cambridge, UK wyrzykow@ast.cam.ac.uk}
\altaffiltext{47}
{Vintage Lane Observatory, Blenheim, New Zealand, whallen@xtra.co.nz}
\altaffiltext{48}
{Auckland Observatory, Auckland, New Zealand, gwchristie@christie.org.nz}
\altaffiltext{49}
{Dept.\ of Physics, Texas A\&M University, College Station, TX, USA, depoy@physics.tamu.edu}
\altaffiltext{50}
{Department of Physics, Institute for Basic Science Research, Chungbuk National University, Chongju 361-763, Korea; cheongho@astroph.chungbuk.ac.kr}
\altaffiltext{51}
{Korea Astronomy and Space Science Institute, Daejon 305-348, Korea; leecu,bgpark@kasi.re.kr}
\altaffiltext{52}
{Farm Cove Observatory, Centre for Backyard Astrophysics, Pakuranga, Auckland New Zealand; farmcoveobs@xtra.co.nz}
\altaffiltext{53}
{Campo Catino Observatory, P.O. Box 03016,  GUARCINO (FR), ITALY; francomallia@campocatinobservatory.org}
\altaffiltext{54}
{Bronberg Observatory, Centre for Backyard Astrophysics Pretoria, South Africa, lagmonar@nmisa.org}
\altaffiltext{55}
{AUT University, Auckland, New Zealand. tim.natusch@aut.ac.nz}
\altaffiltext{56}
{Southern Stars Observatory, IAU/MPC Code 930, Tahiti French Polynesia; santallo@southernstars-observatory.org}


\begin{abstract}
We present the discovery of a Neptune-mass planet OGLE-2007-BLG-368Lb with a planet-star mass 
ratio of $q=[9.5 \pm 2.1] \times 10^{-5}$ via gravitational microlensing.
The planetary deviation was detected in real-time thanks to the high cadence
of the MOA survey, real-time light curve  monitoring and intensive follow-up observations.
A Bayesian analysis returns the stellar mass and distance
at $M_l = 0.64_{-0.26}^{+0.21}$ $M_\sun$  and $D_l = 5.9_{-1.4}^{+0.9}$ kpc, respectively,
so the mass and separation of the planet are
$M_p = 20_{-8}^{+7}$ $M_\oplus$ and $a = 3.3_{-0.8}^{+1.4}$ AU, respectively.
This discovery adds another cold Neptune-mass planet to the planetary sample discovered
by microlensing, which now comprise four cold Neptune/Super-Earths, five gas giant planets,
and another sub-Saturn mass planet whose nature is unclear.
The discovery of these ten cold exoplanets by the microlensing
method implies that the mass ratio function of cold exoplanets scales as
$dN_{\rm pl}/d\log q  \propto q^{-0.7\pm 0.2}$
with a 95\% confidence level upper limit of
$n < -0.35$ (where $dN_{\rm pl}/d\log q \propto q^n$). As microlensing is most sensitive 
to planets beyond the snow-line, this implies that Neptune-mass planets are 
at least three times more common than Jupiters in this region at the 95\% 
confidence level.

\end{abstract}


\section{Introduction}
\label{sec:introduction}
Since the first discovery of exoplanets orbiting main sequence stars in 1995 
(\citealt{may95,mar05}), more than 300 exoplanets have been discovered via the radial 
velocity method (\citealt{may04}) and more than 50 have been detected
via their transits (\citealt{uda04,kon05}). Several planetary candidates have also been
detected via direct imaging \citep{Marois08,Lagrange09},
and astrometry \citep{Pravdo09}. Here, we report the tenth exoplanet discovery by the
microlensing method, which is another example of a cold, Neptune-mass planet
discovered. Although the radial velocity and transit discoveries are 
more numerous, microlensing is uniquely sensitive to these cold Neptunes, and
the microlensing results to date indicate that this class of planets may be the most common type 
of exoplanet yet discovered.

\cite{Liebes64} and \cite{mao91} first proposed exoplanet searches via gravitational microlensing.
The planet's gravity induces small caustics, which
can generate small deviations in standard \citep{pac86} single lens microlensing light curves.
Compared to other techniques, microlensing is sensitive to smaller planets, down to
an Earth mass \citep{bennett96}, and in wider orbits of 1-6 AU.
Because microlensing observability does not depend on the light from the lens host star, 
it is sensitive to planets orbiting faint host stars like M-dwarfs and even brown dwarfs. 
Furthermore, it is sensitive to distant 
host stars at several kpc from the Sun, which allows the Galactic distribution of planetary
systems to be studied.

In 2003, the gravitational microlensing method yielded its first definitive
exoplanet discovery
(\citealt{bon04}). So far 8 planetary systems with 9 planets have been found by 
this technique (\citealt{uda05,bea06,gou06,gau08,bennett08,dong09b,Janczak10}),
which have very distinct properties from those detected by other techniques.
\cite{bea06} found a $\sim 5.5$ Earth-mass planet, 
which was the lowest-mass planet detected at that time.
This detection and the discovery of a slightly more massive planet by
\citet{gou06} demonstrated 
that microlensing is well suited to detecting low-mass planets at orbital distances
that are currently beyond the reach of other methods.
At the time of the discovery of these two cold Neptune-mass planets (hereafter ``Neptunes") or 
``Super Earths", two Jovian planets had also been found.
These discoveries indicate that
cold Neptune in orbits beyond the ``snow-line'' (\citealt{ida04,lau04,ken06}) 
around late-type stars, are significantly more common than gas giants
with frequency of $\ge$16\% at 90\% confidence (\citealt{gou06}), which
is consistent with theoretical simulations (\citealt{ida04}) based on the 
core accretion model. On the other hand, microlensing has also revealed
the most massive M-dwarf planetary companion \citep{uda05,dong09a},
which would likely be difficult to form by core accretion \citep{lau04}.
\cite{gau08} discovered a system with a Jupiter and a Saturn 
orbiting an M dwarf in a configuration very similar to that of our solar
system. Remarkably, this event yielded a direct measurement of the masses of
the planets and the host star, that was confirmed by direct observation of
the host star. This system (OGLE-2006-BLG-109Lb,c) is the only known
multi-planet system with measured masses for the star and planets (aside
from our own Solar System). The light curve of this event
also yielded information about the orbit of the Saturn-mass
planet that confirms that this system is similar to ours \citep{bennett09ogle109}.
A planet was also found to orbit a very low-mass host star or brown dwarf
\citep{bennett08}, and this planet was also the lowest mass exoplanet known
at the time of its discovery.

Here we report the discovery of another Neptune-mass exoplanet in the microlensing 
event OGLE-2007-BLG-368.  We describe the datasets in Section \ref{sec:data}.
The light curve modeling and uncertainty of the parameters are presented in 
Section \ref{sec:lightcurve}, and the physical characterization of the lens system 
is considered in Section \ref{sec:LensMass}. In Section~\ref{sec:mass_func}, we discuss
the implications of microlensing planet discoveries for the exoplanet mass
function. The discussion and conclusions 
are given in Section \ref{sec:conclusion}.

\section{Observations}
\label{sec:data}
The Optical Gravitational Lensing Experiment (OGLE) \citep{uda03} and 
Microlensing Observations in Astrophysics (MOA) \citep{bon01,sume03} are 
conducting microlensing surveys toward the Galactic bulge to find exoplanets.  
From 2002 to 2008, the OGLE-III survey discovered about 600
microlensing events every year by using 1.3m Warsaw telescope with a 0.34 deg$^2$
field-of-view (FOV) mosaic CCD camera with its Early Warning System 
(EWS, \citealt{uda03}). The data have been analyzed in real time and
all kind of deviations from the usual single lens light curves,
including planetary anomalies,
have been detected by the EEWS system \citep{uda03}.
The second phase of MOA, MOA-II, carries out a very high cadence
photometric survey of the Galactic bulge with the 1.8m MOA-II telescope with a 2.2 deg$^2$ 
FOV CCD camera. In 2007, 4.5 deg$^2$ of the central Galactic bulge were observed 
every 10 min, and additional 45 deg$^2$ were observed with a 50 min cadence. This strategy 
enables the detection in real-time of planetary deviations in any of the $\sim 500$ microlensing 
events seen by MOA every year. (Starting in 2010, the new 1.4 deg$^2$ OGLE-IV camera will 
enable OGLE to follow a similar strategy of high cadence monitoring for planetary signals.)

The microlensing event OGLE-2007-BLG-368 was discovered at 
(R.A., Dec.)(2000)= (17:56:25.96, -32:14:14.7) [($l,b$) = (358.3$^\circ$, -3.7$^\circ$)]
and alerted by the OGLE EWS system \citep{uda03} on 2007 July 10, and independently 
detected by MOA and alerted as MOA-2007-BLG-308 on 2007 July 12.

\begin{figure}
\epsscale{0.5}
\includegraphics[angle=-90,scale=0.7,keepaspectratio]{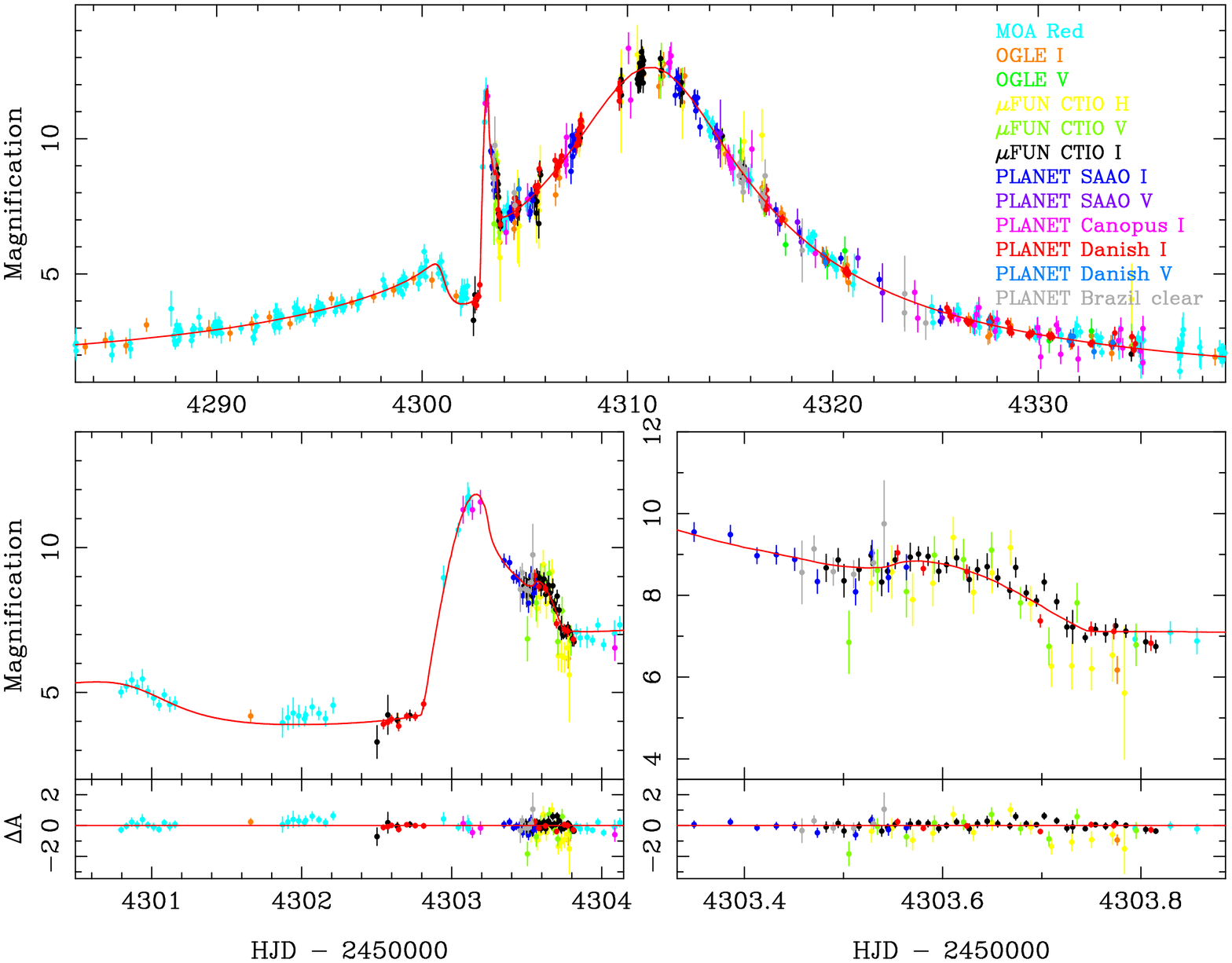}
\caption{
The light curve of OGLE-2007-BLG-368 over the whole event (top-panel), around the planetary 
deviation (lower-left panel) and  the second caustic crossing (lower-right panel) with the 
residual from the best fit model.
The red lines indicate the best fit xallarap model with the Kepler constraint
(see Section \ref{sec:xallarap-kepler}).
Here the light curves of $\mu$FUN CTIO $I$, $H$ and PLANET Brazil are binned by 0.01, 0.01 and 0.02 days,
respectively, for clarity. Note that the fittings were carried for un-binned light curves.
\label{fig:lc}}
\end{figure}

Around UT 12:00 20 July (JD= 2454302), MOA observed a series of 9 points that are all below the
point lens lightcurve, and these are confirmed by a single OGLE point,
with higher precision.  See Figure \ref{fig:lc}. The prompt informative data release to the 
scientific community allowed the SIGNALMEN anomaly detector 
\citep{Dominik2007} (now an integral part of the Automated Robotic Terrestrial 
Exoplanet Microlensing Search (ARTEMiS) system \citealt{Dominik2008})
to detect a light curve anomaly that was passed on to 1--3 members of each of the
major microlensing collaborations, such as the PLANET,
$\mu$FUN, RoboNet, OGLE and MOA at UT 19:32 20 July (JD=2454302.314), 
that this was a possible planetary anomaly. 
Given the intensity of microlensing decision-making and the incompleteness of 
the information flow, this distribution proved only partially adequate and 
failed to reach the MOA internal alert system.
Based on this alert, the $\mu$FUN SMARTS (CTIO) telescope began obtaining 
data just 5 hours later, shortly after dusk in Chile, after which the PLANET 
Danish (La Silla) and Canopus (Tasmania) telescopes also began observations. 
Although MOA observer did not receive this alert, its high-cadence survey 
enabled good coverage of a steep rise due to the caustic entrance in the 
light curve the next night, which triggered MOA's real-time anomaly alert system
to circulate an alert, calling for the firm detection of the anomaly, based on 
its own data at UT 15:58 21 July (JD=2454303.16528).  
Here the real-time anomaly alert system adds new data points on the lightcurves
within 5 min after exposures to search for deviations from the single lens lightcurve.
This continuous early coverage proved crucial for the interpretation of the planetary anomaly. 
See Figure \ref{fig:lc}.
Prompted by these anomaly alerts, MOA-II, OGLE-III and other telescopes from 
PLANET and $\mu$FUN began intensive follow-up observations, which densely covered
the second peak, due to the caustic exit, and less densely for about 50 days.

Twelve light curves were collected by 7 telescopes.
MOA-II 1.8m (Mt. John, New Zealand) obtained $1577$ images in the MOA-Red wide band, which 
corresponds roughly to a combined $I+R$ filter.
OGLE-III 1.3m (Las Campanas, Chile) obtained $12$ images in the $V$ band and $733$ in $I$.
$\mu$FUN SMARTS 1.3m (CTIO, Chile) obtained $22$ images in $V$, $137$ in $I$ and $128$ in $H$.
PLANET SAAO 1m (SAAO, South Africa) obtained $9$ images in $V$ and $60$ in $I$.
PLANET Canopus 1m (Tasmania, Australia) obtained $50$ images in $I$.
PLANET Danish 1.54m  (La Silla, Chile) obtained $20$ images in $V$ and $129$ in $I$.
PLANET OPD/LNA 0.6m (Brazil) obtained $122$ unfiltered images.


The photometry of this event was much more difficult than the photometry of
most microlensing events due to a much brighter star located approximately
$1.1^{\prime\prime}$ to the NW of the source star. This causes very severe problems
with standard PSF-fitting photometry approaches, such as DOPHOT \citep{dophot},
so the only viable approach is the Difference Image Analysis (DIA) method
\citep{tom96,ala98,ala00}.
The images were reduced by three different implementations of DIA photometry.
OGLE V and I and CTIO I images were reduced by the standard OGLE DIA pipeline \citep{uda03}.
Other images were reduced by both the MOA DIA pipeline \citep{bon01} and a version of 
pySIS (v3.0) \citep{Albrow2009}, partly based on ISIS \citep{ala98}, 
but using a numerical kernel \citep{Bramich2008}.
In the MOA DIA pipeline, point spread function (PSF) photometry was performed on 
the difference images with various reference images and PSF fitting radii. 

The resulting MOA DIA light curves, the pySIS light curves, and OGLE DIA light curves 
were compared and the best one was selected in each data set as follows.
First, the planetary deviation at HJD-245000 = 4300--4305 was removed from each 
light curve, and these planet-free light curves were fitted with a 
single lens model with xallarap (binary orbital motion of the source). (Details are 
discussed in Section \ref{sec:lightcurve}). The photometric reduction yielding 
the smallest variance from the best model in these planet-free fits were selected 
to use for further analysis. For each data set, the error bars are rescaled so that
$\chi^2/{(\rm d.o.f.}) \approx 1$ in the planet-free single-lens fit.
For CTIO $V$ and $H$ which have very few data points unaffected by the planetary deviation, 
this same procedure was followed including the planetary deviation 
using the best fit planetary model to all the data sets.

Figure \ref{fig:lc} shows the light curves of this event around the peak and the planetary deviation.

\section{Light Curve Modeling}
\label{sec:lightcurve}

\begin{figure}
\epsscale{0.5}
\includegraphics[angle=-90,scale=0.65,keepaspectratio]{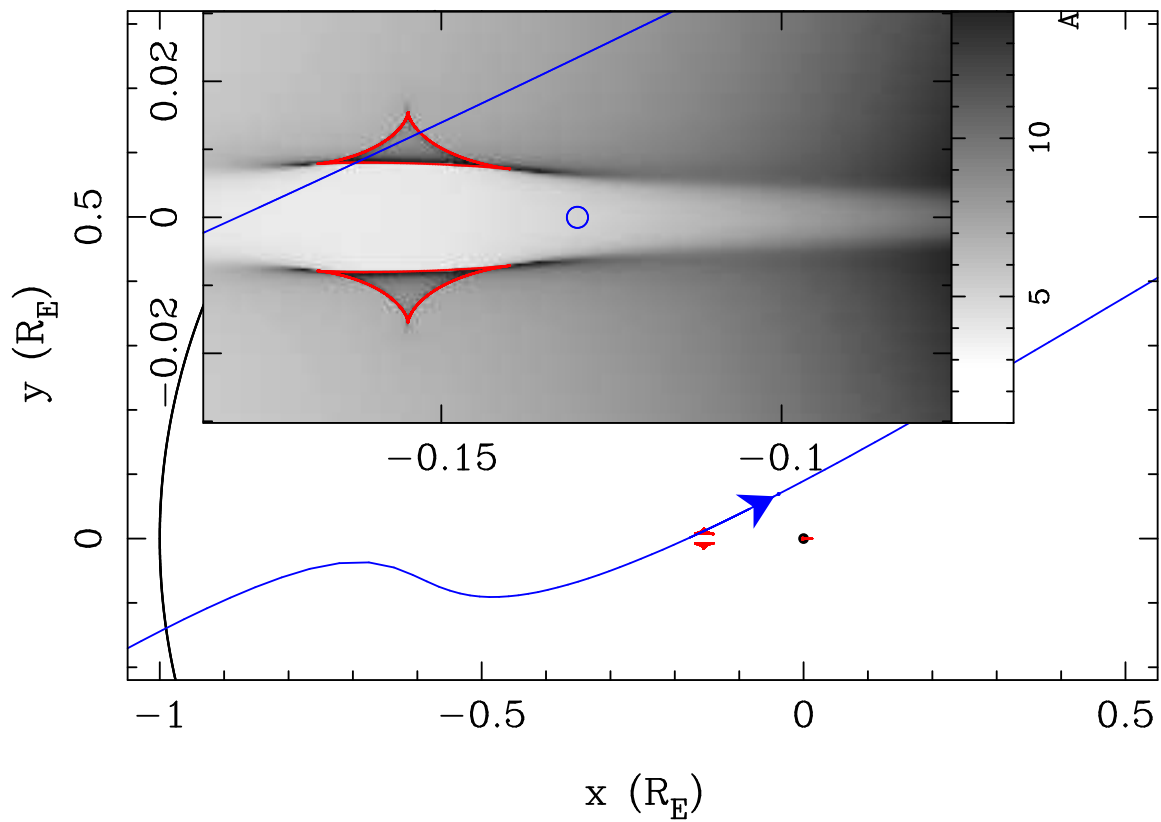}
\caption{
The caustics (red lines) and critical curves (black lines) of OGLE-2007-BLG-368 
for the best model fitting $\epsilon$ with the Kepler constraint (see Section \ref{sec:xallarap-kepler}).
The blue line represents the trajectory of the source.
The inset shows the zoom around the planetary caustic crossing, where
the gray scale indicates the magnification pattern.
The circle in the inset represents the best fit source size.
\label{fig:caustics}}
\end{figure}

The negative deviation that triggered the initial alert is characteristic of 
``minor image perturbations", in which the image inside the Einstein ring is 
perturbed by a planet, and therefore a planet is inside the Einstein ring. 
In this case, two triangular caustics appear near the central caustic, 
on the opposite side of the planet, as
as shown in Figure \ref{fig:caustics}.
The Danish (La Silla) data show a caustic entrance just prior to their last point,
and the MOA and Canopus data confirm this entrance and trace its rise.
From these data alone it is clear that the source has passed into the
``depression" between the two triangular caustics and then passed over
one of the two parallel caustic walls that bound this depressed region.
See Figure \ref{fig:caustics}.  The subsequent data over the next day trace the path
through a triangular caustic. A blind search of parameter space, in which 
$\chi^2$ minimizations were done with various initial parameters,
confirms that this is the only viable topology.  

In addition to the three single lens model parameters, the time of peak magnification 
$t_0$, Einstein radius crossing time $t_{\rm E}$ and the minimum impact parameter $u_{0}$,
the standard binary lens model has four more parameters, the planet-host mass ratio $q$, 
projected separation $d$,
the angle of the source trajectory relative to the binary lens axis $\alpha$, and
source radius relative to the Einstein radius $\rho=\theta_{\rm *}/\theta_{\rm E}$,
or the source radius crossing time $t_*=\rho t_{\rm E}$.
Note that $\rho$ can be used to estimate angular Einstein radius $\theta_{\rm E}$ by using
the source angular radius $\theta_{\rm *}$ which can be estimated from its color and
apparent magnitude \citep{yoo04}.

A simple heuristic argument can be given to derive $q$, $d$ and $\alpha$
 from the gross characteristics of the light curve \citep{GouldLoeb1992,Gaudi1997}.
From the point-lens part of the light curve with the planetary
perturbation excluded, we robustly find $t_0\simeq2454311$ JD, $t_{\rm E}\simeq55$ days and $u_0\simeq0.08$.
The time and duration of the planetary deviation is $t_{\rm d}\simeq2454303$ JD and $\Delta t\simeq1$ day,
where we adopt the duration of the negative deviation relative to the single lens model.
By using these planet-model independent values, the position of the planet can be estimated as
$d_{-}=(\sqrt{u_{\rm d}^2+4}-u_{\rm d})/2=0.92$, where 
$u_{\rm d}=\sqrt{\tau(t_{\rm d})^2+u_0^2}=0.166$ and $\tau(t_{\rm d})=(t_{\rm d}-t_0)/t_{\rm E}$.
The angle of the source trajectory relative to the binary lens axis, $\alpha$, can be given by 
$\sin\alpha=u_0/u_{\rm d}=0.48$, therefore $\alpha=0.5$ rad. 
The separation of the two triangular caustics is given by $d_{\rm caus}=2(\gamma-1)^{1/2}$
in the unit of the planet angular Einstein radius $\theta_{p}=q^{1/2}\theta_{\rm E}$ 
\citep{Schneider1992}, where $\gamma=d_{-}^{-2}$ is the shear.
The duration required to pass the ``depression" between the two triangular caustics is
given by $\Delta t=2(\gamma-1)^{1/2}q^{1/2}(\csc\alpha) t_{\rm E}$.
Therefore, we find that the planet has the sub-Saturn mass ratio, $q\sim1\times 10^{-4}$.
From the light curve around JD$=2454303$, we can roughly find the time it takes the source radius
to cross the caustic $t_{\rm cross}\simeq 0.23$ days.
Therefore the source radius crossing time can be estimated as $t_* = t_{\rm cross} \sin\alpha\sim0.1$ day.
These first order estimates of the planetary modeling are very robust.
The actual light curve modeling will investigate several higher order effects and 
possible systematics, but all within the context of the topology defined by 
Figure \ref{fig:caustics}, and the main conclusions remain robust.

The light curve modeling was done by two independent codes.
One uses the hybrid point-source, individual image ray-shooting 
method of \cite{bennett09}, which has been developed
from the first completely general finite-source binary lens calculations of
\cite{bennett96}. The other uses the same basic strategy,
but was independently written by MOA. 
The best fit binary lens model was found by the Markov Chain Monte Carlo 
(MCMC) method \citep{verde03}.
The Markov chains of preliminary runs were used to derive the optimal directions and 
step sizes for exploring parameter space. The resultant distribution of the 
chains gives us the best fit parameters and their errors.
We use a linear limb-darkening model for the source star 
using the coefficients, $u=0.5250$ for $I$-band, $0.6834$ for $V$,
$0.3434$ for $H$ and $u=0.566$ for MOA-Red which is a mean of $R$ and $I$,
from \cite{Claret2000} for a 
G6 type source star with $T=5750$ K and log$g=4$, which is based 
on the best fit source $V-I$ color (see Section \ref{sec:source}).
The best fit source and blend are plotted in the color magnitude diagram (CMD) 
(Figure \ref{fig:cmd}).
The best fit standard binary lens model has a planetary mass ratio of $q=1.3\times 10^{-4}$
and other parameters as listed in Table \ref{tbl:param}, in which $q$, $d$ 
$\alpha$ and the source radius crossing time, $t_* = \rho t_{\rm E}=0.1$ day
are consistent with the first order estimates given above.

However the overall light curve shows asymmetric residuals about the primary peak, 
which suggests either the microlensing parallax effect \citep{gou92,alc95,smi02} 
by which the Earth's orbital motion distorts the light curve and/or the xallarap effect, 
which is a similar distortion caused by the orbital motion of a binary source \citep{Griest92,han97}.
Therefore, we compare the data to models that included parallax and xallarap.

\subsection{Microlensing Parallax}
\label{sec:parallax}
The parallax effect is represented by two additional parameters, an amplitude 
$\pi_{\rm E}=\pi_{\rm rel}/\theta_{\rm E}$, i.e., the lens-source relative parallax
$\pi_{\rm rel}=(\pi_{\rm l} - \pi_{\rm s})$ in unit of the angular Einstein radius 
$\theta_{\rm E}=R_{\rm E}/D_l$, and a direction of the relative source-lens proper 
motion relative to North toward East $\phi_{\rm E}$, where $D_l$ is the distance to 
the lens. As shown in Table \ref{tbl:param}, the best fit parallax model improves
$\chi^2$ by $\Delta \chi^2=298$ relative to the best standard binary model.
If this parallax model were the correct model, we could derive the lens mass
$M = {\theta_{\rm E} / (\kappa \pi_{\rm E})} = 0.040 \pm 0.005\, M_\sun$, 
and distance $D_l = {{\rm AU}/ \pi_l \rm}=867 \pm 93\, \rm pc$, 
for this model of the lens \citep{gou92}.
Here $\kappa=4G/c^2AU = 8.144$mas\,$M_\sun^{-1}$ (milli-arcsec per solar mass) and 
we have assumed the source distance $D_s={\rm AU}/\pi_{\rm s}=8.0 \pm 1.4$ kpc
where the error is based on 17\% standard deviation in the Galactic bar model \citep{han03}.
This model implies that the lens is a nearby brown dwarf. 
However, as shown in the next section, the 
best xallarap model yields a significantly better $\chi^2$, with an
improvement by $\Delta\chi^2 = 89.4$. Furthermore,
if the signal were due to parallax, we should have found the best xallarap model
with the same (R.A.$_\xi$, Dec.$_\xi$) values as the celestial coordinates of 
the source as seen from the Earth 
(RA, Dec.)= (269.1$^\circ$, -32.2$^\circ$)
when its period of the source orbital motion, eccentricity, and perihelion,
celestial pole are fixed as the values of Earth's orbit. However, we obtained 
the best model with
(R.A.$_\xi$,  Dec.$_\xi$ )=(309.4$^\circ$, -24.0$^\circ$) $\pm$ (0.5$^\circ$, 0.2$^\circ$),  
which is inconsistent with the expected values for parallax. 
We conclude that this distortion is not likely due solely to parallax.

\subsection{Xallarap}
\label{sec:xallarap}

If the orbit is assumed circular, and the companion
assumed to generate negligible flux compared to the source,
the xallarap effect can be represented
by five additional parameters, an amplitude,
$\xi_{\rm E}=a_s/\hat{r}_{\rm E}$, that is the semi-major axis of the source's orbit, $a_s$,
in the unit of the Einstein radius projected on the source plane,
$\hat{r}_{\rm E}= R_{\rm E}D_s/D_l$, 
the direction of the relative source-lens proper motion, $\phi_\xi$, the direction of observer 
relative to the source orbital axis, R.A.$_\xi$ and Dec.$_\xi$, orbital period, $P_\xi$.
For an elliptical orbit, two additional parameters are required, the orbital eccentricity,
$\epsilon$ and time of perihelion, $t_{\rm peri}$.

The best xallarap model models, with $\epsilon$ fixed at $\epsilon =0$
and with $\epsilon$ as a free parameter, improved $\chi^2$ by 
$\Delta \chi^2= 74$ and $89$, respectively, relative to the the best parallax model. 
The best fit parameters are listed in Table \ref{tbl:param}.
We also fitted models with a bright binary companion, but in every case, the dark binary 
companion provided the best fit.
Therefore we keep only models having companions
with negligible flux compared to the source in the following analysis, which would
be appropriate for a white dwarf or M dwarf companion.

\begin{figure}
\epsscale{0.5}
\includegraphics[angle=-90,scale=0.6,keepaspectratio]{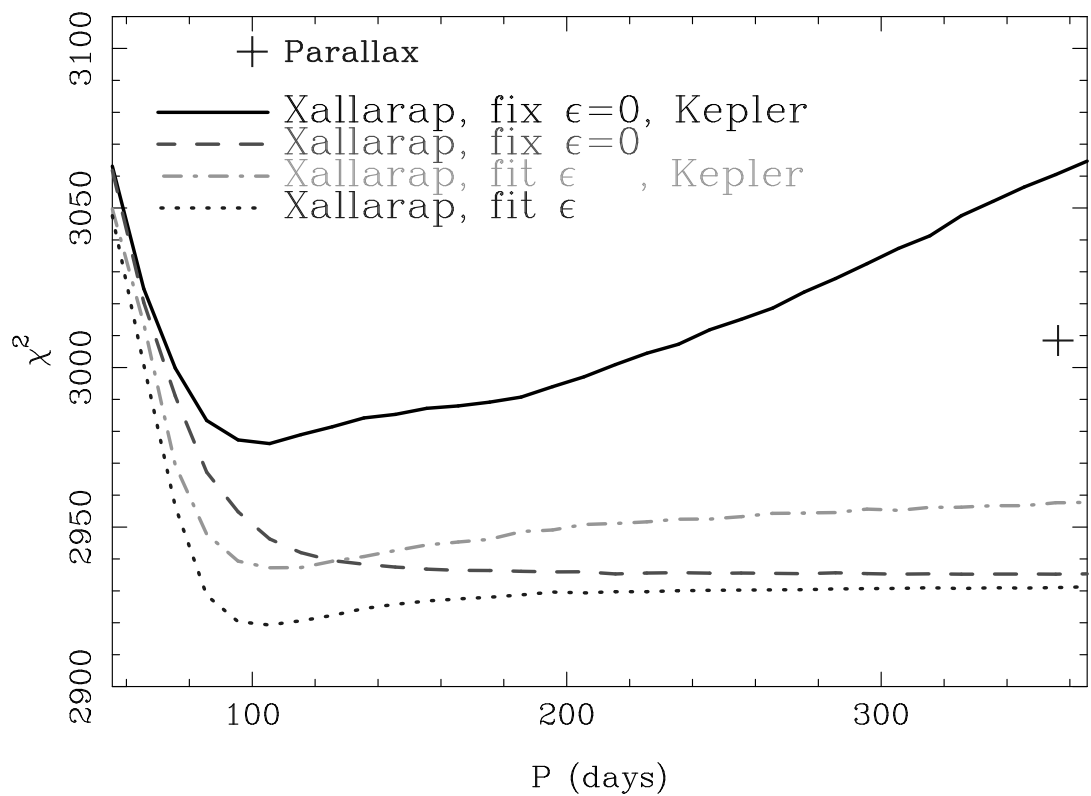}
\caption{
The $\chi^2$ of the best xallarap model as a function of the orbital period of the source 
star and its companion. 
The solid and dashed lines indicate the model with fixed orbital 
eccentricity $\epsilon=0$ with and without the Kepler constraint, respectively.  
The dot-dashed and dotted lines indicate the model allowing a free-fit of $\epsilon$ 
subject and not subject to the Kepler constraint, respectively.  
The best parallax model is plotted as a "+" for comparison.
\label{fig:chi2Px}}
\end{figure}

In Figure \ref{fig:chi2Px} we show the $\chi^2$ of the best xallarap models as a 
function of $P_\xi$ with orbital eccentricity fixed at $\epsilon=0$ and 
fitting for $\epsilon$.
One can see that xallarap models are significantly better than the best parallax model, 
and that the xallarap model in which $\epsilon$ is a free parameter is slightly better 
than the model fixing $\epsilon=0$.
For the xallarap models, $\chi^2$ is flat for $P_\xi \ge 150$days, 
in which regime $P_\xi$ and $\xi_{\rm E}$ are strongly degenerate.

Of course, every microlensing event must have a non-zero microlensing parallax, but
the addition of parallax to these xallarap models did not provide a significant
$\chi^2$ improvement. The parallax and xallarap parameters are highly degenerate
and tend to complicate the analysis, so we have excluded parallax from most
of our xallarap models.

\subsection{Xallarap with the Kepler Constraint}
\label{sec:xallarap-kepler}

In Section \ref{sec:xallarap}, the model with the lowest $\chi^2$ is the 
xallarap model with non-zero $\epsilon$. However, this model leads to a 
xallarap amplitude of $\xi_{\rm E}=0.35$, which is larger than would be 
induced by a ``normal'' main-sequence companion, where
$\xi_{\rm E}$ is expressed, making use of Kepler's third law, by
\begin{equation}
\xi_{\rm E}={a_s \over \hat{r}_{\rm E}}
    ={{1\rm AU}\over \hat{r}_{\rm E}} 
    \left(  {M_c \over M_\sun } \right) 
  \left(  { M_\sun \over M_c+M_s }  {P_\xi \over 1 \rm yr} \right)^{2\over3}.
\label{eq:xiE}
\end{equation}
From this equation and parameters for this model, the lower limit of the companion
mass to the source is given by,
\begin{equation}
M_c \ge \xi_{\rm E}^3 \hat{r}_{\rm E}^3  \left( {P_\xi \over 1 \rm yr} \right)^{-2} \sim 50 M_\sun, 
\label{eq:Mc}
\end{equation}
which would imply a black hole companion, since a $50\,M_\sun$ star would 
exceed our upper limit on the brightness of a companion to the source by
more than 5 magnitudes.

Black holes are quite rare compared to stars, so we should consider the prospect
of a more normal stellar companion.
We can use Kepler's third law, the projected Einstein radius, 
$\hat{r}_{\rm E}$, source mass, $M_s$,
and the source companion mass, $M_{c}$, to constrain
the magnitude of the xallarap vector $\xi_{\rm E}$ \citep{bennett08,dong09b}. From Section \ref{sec:source},
we derive $M_s= 0.9\pm 0.1 M_\sun$  
and assume a white dwarf companion $M_{c}= 1.0 \pm 0.4 M_\sun$
(which would be the most massive dark companion with a plausible a priori probability).
Inserting these masses and other relevant parameters into
Eq. (\ref{eq:xiE}), the maximum allowed $\xi_{\rm E}$ for the best xallarap
models for the circular orbit xallarap and non-zero $\epsilon$ models are given by
\begin{equation}
\xi_{\rm E,max}
=  0.11 \pm 0.04 \,(\epsilon=0)\ \ \ {\rm and} \,\, 0.06 \pm 0.02 \,(\epsilon \,{\rm free}), 
\label{eq:XiEmax}
\end{equation}
where the error is estimated from the errors in $\theta_*$, $M_s$ and $M_{c}$.

Because our best fit values of $\xi_{\rm E}=1.73$ and $0.35$ for the circular orbit
and non-zero $\epsilon$
are much larger than $\xi_{\rm E,max}$ given above, they are inconsistent with our
upper limit on the source companion mass.
To find the best xallarap model that is allowed by Kepler's third law,
we have done MCMC runs with an additional constraint contribution to $\chi^2$ given by
\begin{equation}
\chi^2_{\rm orb} =\Theta( \xi_{\rm E} - \xi_{\rm E,max} ) \left( { \xi_{\rm E,max}- \xi_{\rm E} \over \sigma_{\xi_{\rm E,max}} }  \right)^2.
\label{eq:chi2XiEmax}
\end{equation}
where  $\xi_{\rm E,max}$ is evaluated by Eq. (\ref{eq:xiE}) with parameters in each
step of the MCMC and fixed values of $M_s = M_{c} =1 M_\sun$ and
50\% error in $\xi_{\rm E,max}$, which depend only weakly on other parameters.
Here, $\Theta$ is the Heaviside step function.

In Table \ref{tbl:param}, we show the best fit model parameters and $\chi^2$ for
the circular orbit and non-zero $\epsilon$ cases.
In Figure \ref{fig:chi2Px} we show the $\chi^2$ of the best-fit xallarap models
with the Kepler constraint as a function of $P_\xi$.
One can see that if we impose the Kepler constraint, the xallarap solution with 
$\epsilon$ free is better than the case of fixed $\epsilon=0$ by $\Delta \chi^2=39$.
Although this $\chi^2$ is worse, by $\Delta \chi^2=18$, than the model without the 
Kepler constraint,
this is the best model that is allowed for a plausible companion mass.

\section{The Errors in Parameters with Systematics}
\label{sec:systematics}

          
We have investigated second order effects to explain the clear
asymmetry about the peak in Section \ref{sec:parallax},
\ref{sec:xallarap} and \ref{sec:xallarap-kepler}.
We also searched for models with an additional mass besides the lens star
and planet, but neither an additional stellar or planetary companion to
the lens star could account for observed light curve asymmetry.

\begin{figure}
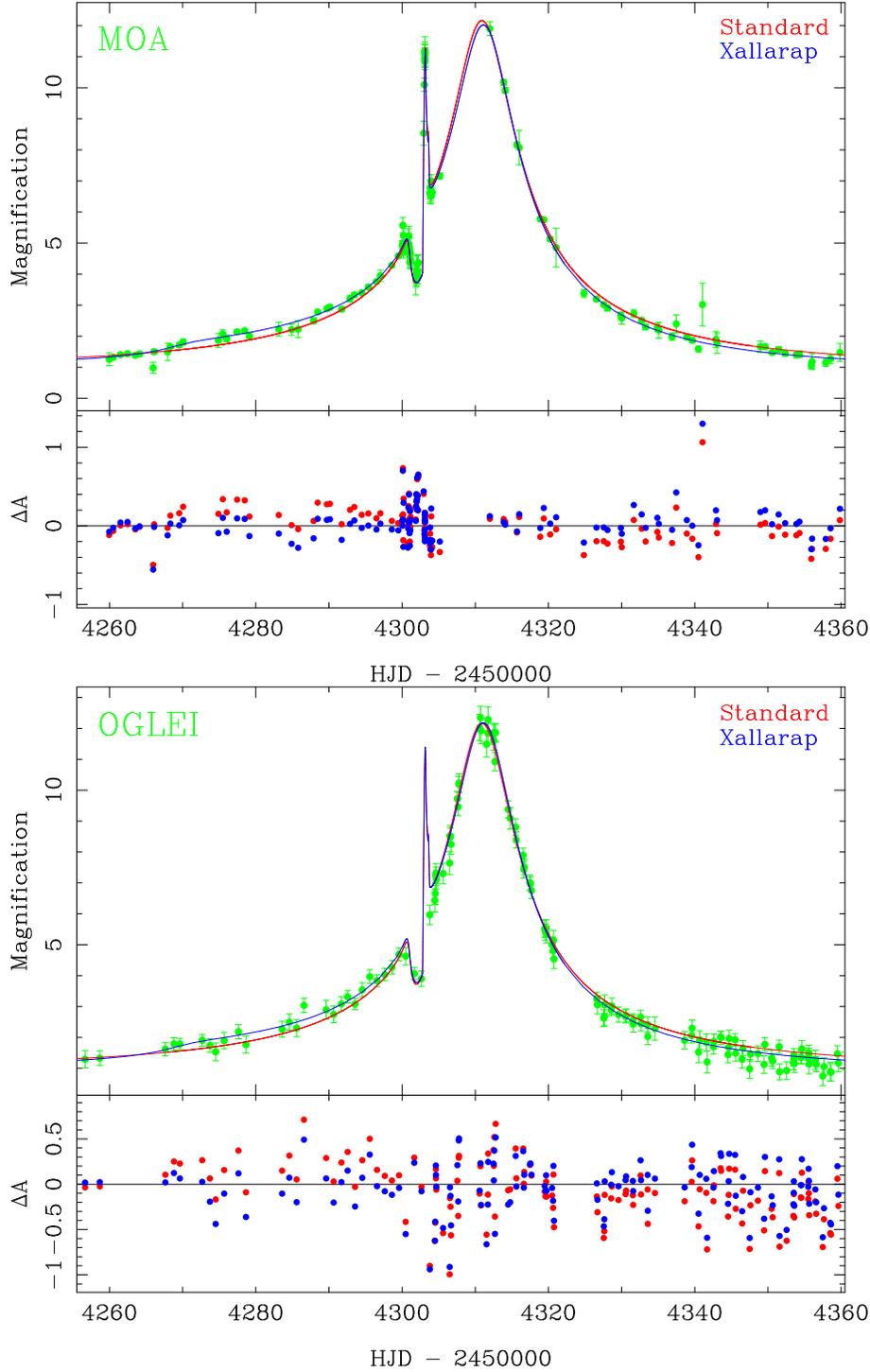

\epsscale{0.5}
\includegraphics[angle=-90,scale=0.5,keepaspectratio]{f5a.eps}
\includegraphics[angle=-90,scale=0.5,keepaspectratio]{f5b.eps}
\caption{
The light curves of MOA Red (top panel) and OGLE I (bottom panel) with the best 
standard (red line and residual) and xallarap models with $\epsilon$ being fit 
subject to the Kepler constraint (blue line and residual). Here MOA data are binned 
by 1 day outside of the planetary signal at JD-2450000=4300-4304. We can see a similar 
asymmetry in both light curves is well fitted by the xallarap model in both cases.
\label{fig:residual}}
\end{figure}

\begin{figure}
\epsscale{0.5}
\includegraphics[angle=0,scale=0.6,keepaspectratio]{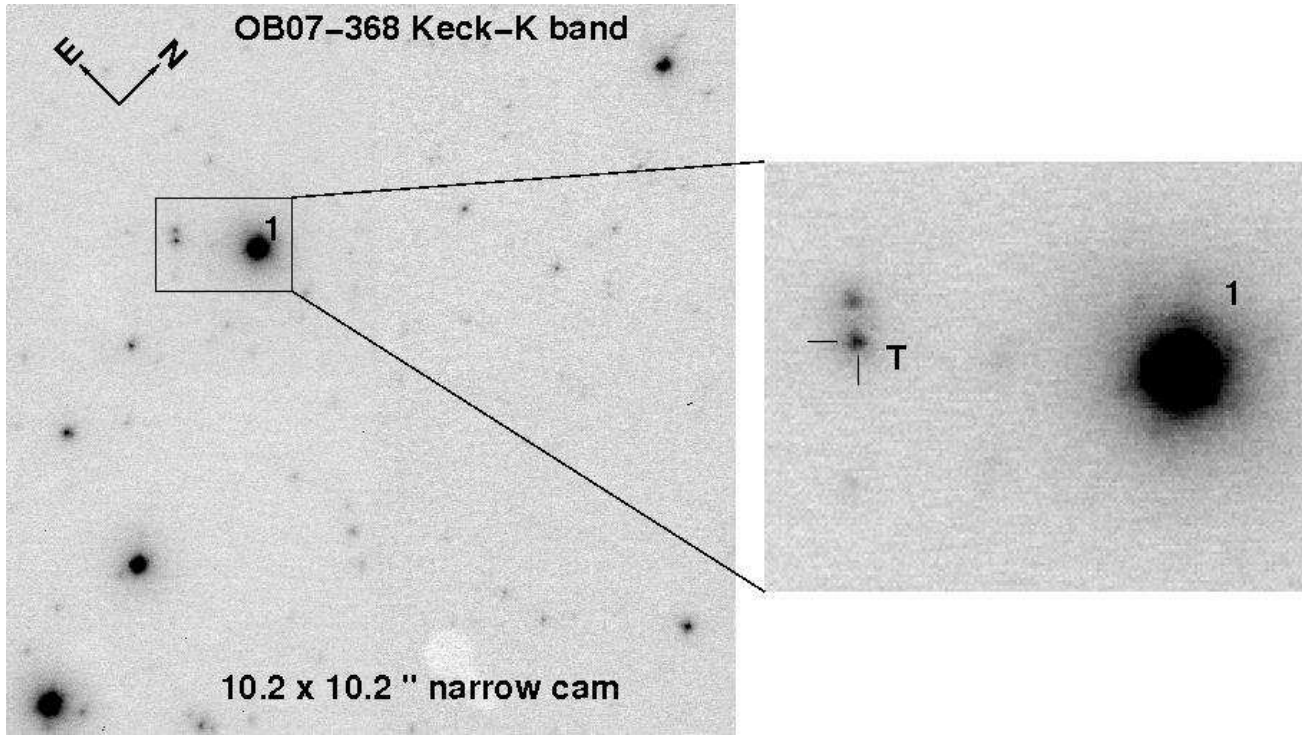}
\caption{
K-band Keck AO narrow camera image of OGLE-2007-BLG-368. 
The magnified source position on the OGLE difference image is marked as ``T'' with 
error of $\sim$2.5 pixels (25mas) in $K$-band,
where the coordinate is aligned by using the 5 brightest stars in the K-band image.
This object is almost certainly the source, the lens and/or their companion. 
Blending with random interloper is unlikely with this stellar density (see Section \ref{sec:keck}).
The bright RCG star marked as ``1'' is $1.1''$ way from the source, whose PSF tail
covers the source on OGLE image with typical seeing  of $1.2''$.
\label{fig:keck}}
\end{figure}

We are sure that there is an asymmetry in the light curve, because we see 
qualitatively similar trends in both MOA and OGLE light curves as shown in
Figure \ref{fig:residual}, which have different typical seeing and were 
reduced by independent pipelines. However we are not fully confident that 
this xallarap amplitude is correct because of the unphysically large 
$\xi_{\rm E}$ and an additional factor: there is a bright red clump 
giant (RCG) star with $(V-I,I)$=(2.01,15.56) at the North-East corner of 
the Keck AO image (see details in Section \ref{sec:keck}) 
in Figure \ref{fig:keck}, which is only 1.1$''$ away from the source.
The wing of the giant star PSF interferes with photometering the source on the 
OGLE images, with typical seeing of $\sim 1.2''$, and even worse in the MOA images, 
with typical seeing of $\sim 2.0''$. 
The differential atmospheric extinction and refraction may cause systematic 
asymmetry on the light curve. Here the differential refraction causes the 
positional shift of the target on the sky, which generates residuals on the 
subtracted images in DIA. These effects depend on the color of stars and airmass.
The mean airmass changes slowly during the event because the mean elevation 
of the target changes in season. They can generally be reduced by choosing 
the reference stars with the similar color as the target for aligning the image 
coordinates and solving the kernel in DIA. However, the effects due to the blending 
star with different color from the target are hard to remove. Especially, subtle 
effects on the bright blending star can cause significant effects on the faint target.
So the photometry of this event is challenging. 
We tested the modeling with data points taken at airmass $> 1.3$ removed, 
but this does not make any significant difference. When we model by first removing 
either the OGLE or MOA dataset, the results are qualitatively unchanged.

As argued in Section \ref{sec:lightcurve}, the planetary deviation is clearly 
detected, and the planet parameters can be estimated robustly by simple inspection.
Although our analysis of the parallax and xallarap fits indicate the presence
of unrecognized systematic errors in the data, these errors do not affect these
basic inferences about the planet itself. Therefore, we are only interested 
in robustly estimating the parallax or xallarap parameters to the extent that 
they can provide additional information about the lens. However, having discarded
the parallax model for the asymmetry, the xallarap parameters themselves provide
no new information about the lens, and are therefore of no intrinsic interest to us.
We therefore do not further investigate the cause of the systematics in the light 
curve, and instead seek only to determine the (relatively minor) extent to which
these systematics affect our precise determination of the planetary parameters.  
To do this, we consider the standard and all the possible xallarap models shown 
in Table \ref{tbl:param} as viable, and take the differences of the parameters 
as the size of the systematic errors.
We take parameters of the xallarap model with non-zero $\epsilon$ and the
Kepler constraint (indicated as ${\rm Xallarap}^K$ in Table \ref{tbl:param}), as the median.
The resultant systematic errors are listed in Table \ref{tbl:param} and
$I_{\rm s,OGLE}=19.51 \pm 0.03$ mag and $I_{\rm b,OGLE}=18.77 \pm 0.02$ mag.
The values and errors in the following sections are estimated by taking
these systematic errors into account.


\section{Source Star Characterization}
\label{sec:source}
We must determine the source star angular radius, $\theta_*$, in order to determine
the angular Einstein radius, $\theta_E$, from the light curve parameters. Since
we do not have infrared light curve data of high enough quality to accurately measure
the source brightness in the infrared \citep{ogle224,bennett09ogle109}, we use
the two-filter method of \citet{yoo04} to determine $\theta_*$.
The values and errors stated in this section are the final 
values including systematic errors, as they are subsequently re-estimated after Section \ref{sec:systematics}. 
However, these are qualitatively the same within the errors as the original results 
used for the initial light curve modeling in  Section \ref{sec:lightcurve}.

\subsection{Extinction Correction}

\begin{figure}
\epsscale{0.5}
\includegraphics[angle=0,scale=0.7,keepaspectratio]{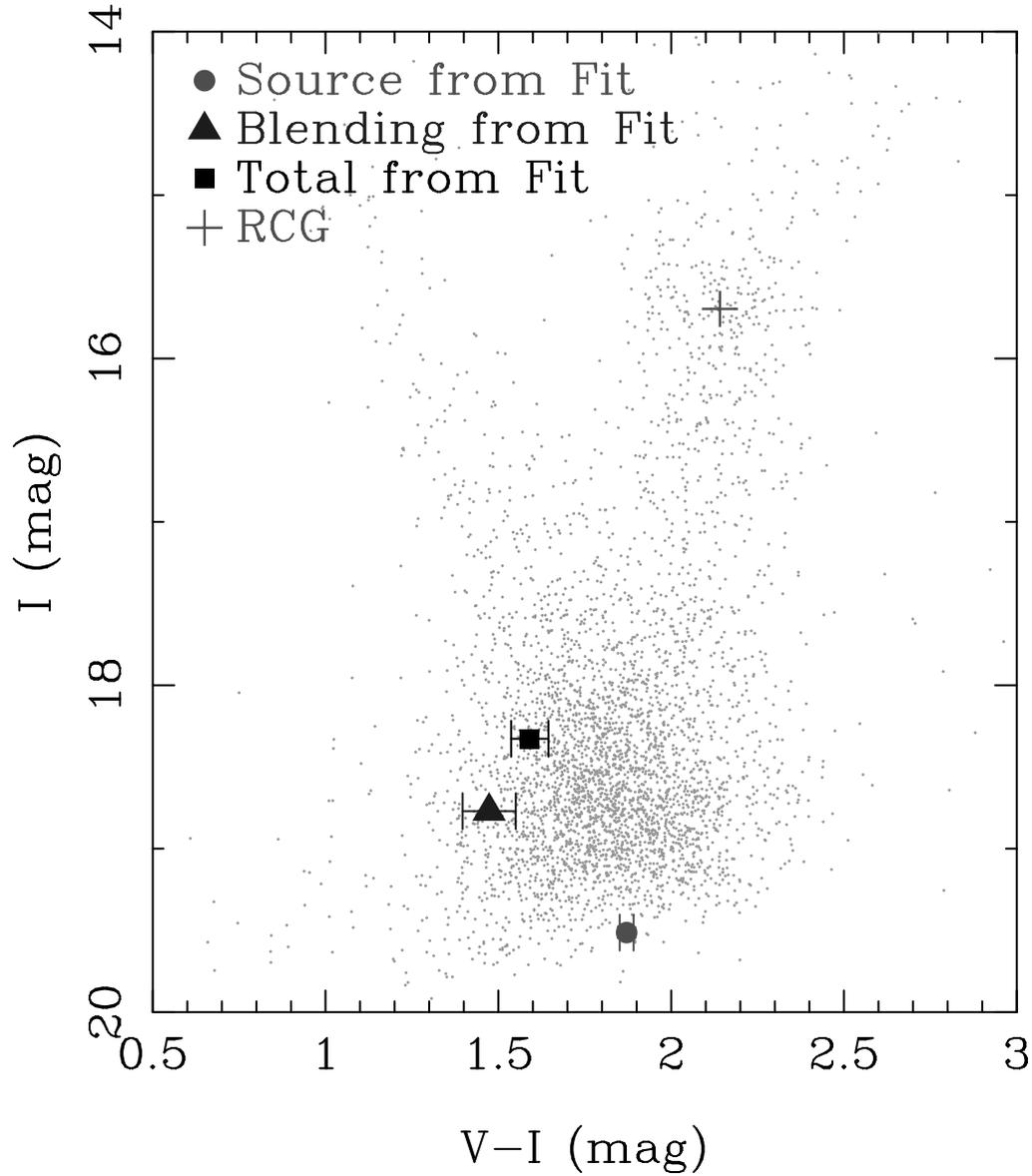}
\caption{
The OGLE ($V$-$I$, $I$) color magnitude diagram around OGLE-2007-BLG-368. The filled 
circle and triangle indicate the source and blended light from the fit, respectively.
The filled square represent the total flux of the source and blend. Here the errors in 
$I$ are too small to be visible. The cross indicates the center of the RCG.
\label{fig:cmd}}
\end{figure}

The $V$ and $I$ magnitudes of the source star from the light curve fit 
(see Section \ref{sec:lightcurve}) must be corrected for the extinction 
and reddening due to the interstellar dust to infer the spectral type of the source.
Because this field is out of the OGLE-II extinction map \citep{sumEX04}, 
we estimate the extinction and reddening to the source
by using RCG, which are known to be an approximate standard candle \citep{sta97,pac98}.
Figure \ref{fig:cmd} shows the calibrated OGLE CMD in 3$\times$3 arcmin$^2$ 
field around the event [($l,b$) = (358.3488$^\circ$, -3.6861$^\circ$)]. 
From this CMD, we find the RCG centroid;
\begin{equation}
      (V-I, I)_{\rm RC,obs} = (2.14, 15.70),
\label{eq:RCGobs}
\end{equation}
where the errors are negligible comparing to the intrinsic error in the RCG centroid,
as described below.

We adopt the intrinsic RCG magnitude $M_{I,\rm RC,0}=-0.25\pm0.05$, 
$M_{V, \rm RC,0}=0.79\pm0.08$ and color $(V-I)_{\rm RC,0}=1.04\pm 0.08$ from \cite{bennett08} 
which is based on \cite{girardi01} and \cite{salaris02}, where the error is 
assigned based on the size of the theoretical corrections to the RCG magnitudes.
Taking account of the bar structure of the Galactic bulge, the offset of the distance 
modulus ($DM$) between the GC that is assumed to be at $8.0\pm0.5$kpc \citep{reid93} 
and the average stars in the field have $\Delta DM=0.00\pm0.05$ \citep{nishiyama05}.
So the dereddened RCG centroid in the field is expected to be
\begin{equation}
      (V-I, I)_{\rm RC,0} = (   1.04, 14.27 ) \pm (  0.08,  0.15 ).
\label{eq:RCGexp}
\end{equation}

Comparing these centroids (Eqs. \ref{eq:RCGobs} and \ref{eq:RCGexp}) , 
we find the average reddening and extinction in this field is
\begin{equation}
(E(V-I), A_I) = ( 1.10,  1.43 ) \pm (  0.08,  0.15 ),
\label{eq:AI}
\end{equation}
where $R_{VI}=A_V/E(V-I)=  2.30$, which corresponds to $R_V=2.64$
\citep{car89}. 
Applying this average extinction to this event,
the source's $(V-I, I)_{\rm s,OGLE}$ from fitting of the well calibrated OGLE $V$ and 
$I$  light curve and the dereddened source magnitude and color  $(V-I,I)_{\rm s,0}$ are
\begin{eqnarray}
(V-I,I)_{\rm s,OGLE}    &=&(  1.85, 19.51 ) \pm ( 0.06, 0.03),\\ 
\label{eq:VIIsOGLE}
(V-I,I)_{\rm s,OGLE,0}      &=&(   0.75, 18.08 ) \pm (  0.10,  0.16), 
\label{eq:VIIsOGLE0}
\end{eqnarray}
Independently, the dereddened source color, $(V-I)_{\rm s,CTIO,0}=0.77 \pm 0.02 \pm 0.08$, 
is estimated by comparing $(V-I)_{\rm RC,0}$, the CTIO RCG color and 
the CTIO source color $(V-I)_{\rm s,CTIO}$ which is given by the 
model-independent regression of CTIO $V$ and $I$ light curves. This is consistent with 
$(V-I)_{\rm s,OGLE,0}$, but more accurate. In the following analysis, we adopt the value:
\begin{equation}
(V-I,I)_{\rm s,0}      =( 0.77 , 18.08 ) \pm ( 0.08 , 0.16). 
\label{eq:VIIs0}
\end{equation}
which implies that the source is a mid-G star in the bulge 
\citep{bessell88}
with mass of $M_s=0.9\pm 0.1 M_\sun$ \citep{schmidt-kaler82}. 
The reddened $(V-I,I)_{\rm s}$ is plotted in Figure \ref{fig:cmd}.

The dereddened blended light from the light curve is
\begin{equation}
(V,I)_{\rm b,0}      =( 17.71 , 17.34 ) \pm (0.18 ,  0.15). 
\label{eq:blend}
\end{equation}
Note that if this blended light is from the lens or companion of the lens, 
these values may be over-corrected for extinction because these objects 
are in front of the source.
Thus, these magnitudes can be used as an upper limit on the combined light of the lens,
any companion of the lens, and the source in the following analysis.

\subsection{Source Star Angular Radius}
\label{sec:sourceradius}
\begin{figure}
\epsscale{0.5}
\includegraphics[angle=0,scale=0.6,keepaspectratio]{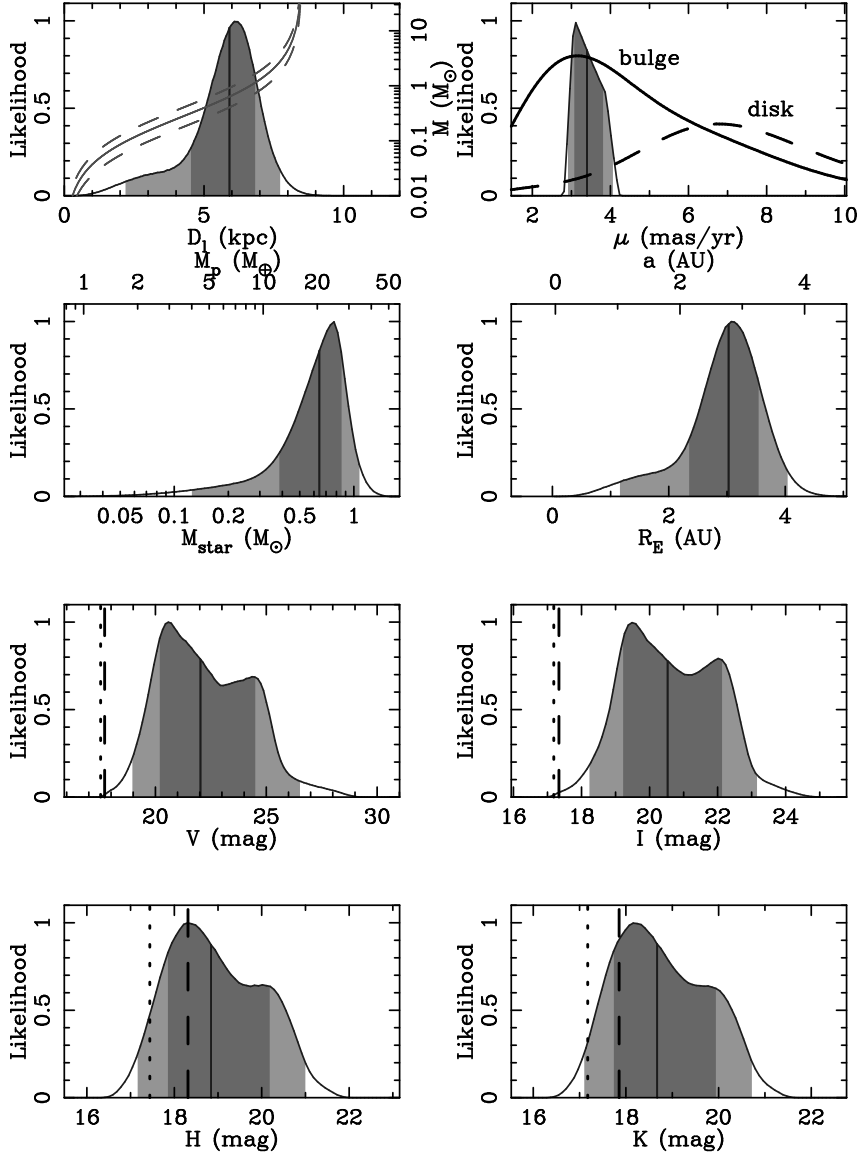}
\caption{
Probability distributions from a Bayesian analysis for the distance, $D_l$, 
transverse velocity, $v_{\rm t}$, mass, $M_{\rm star}$, Einstein radius ($R_{\rm E}$), 
$V$, $I$, $H$ and $K$-band magnitudes of the primary star of the lens system.
The vertical solid lines indicate the median values. The dark and light gray shaded regions 
indicate the 1-$\sigma$ and 2-$\sigma$ limits.
The gray solid and dashed curves in the top-left panel indicate the mass-distance relation of the 
lens from the measurement of $\theta_{\rm E}$ with 1-$\sigma$ errors, respectively, assuming $D_s=8$ kpc.
Note $D_s$ is not fixed in the actual Bayesian analysis.
Thick solid and dashed lines in the top-right panel represent the typical $\mu$ distributions
of the bulge and disk lens populations, respectively.
The vertical dashed and dotted lines in the $V$, $I$, $H$, and $K$-band panels represent observed
upper limit and 1-$\sigma$ error, respectively.
\label{fig:likelihood}}
\end{figure}

Following \cite{yoo04}, the dereddened source color and brightness 
($V$-$K$, $K$)$_{\rm s,0}=(1.69, 17.16)$ are estimated using the observed  
($V$-$I$, $I$)$_{\rm s,0}$ as given by the equation (\ref{eq:VIIs0}) and the 
color-color relation 
\citep{bessell88}.
By using this ($V$-$K$, $K$)$_{\rm s,0}$ and the empirical color/brightness-radius relation by 
\cite{kervella2004}, 
we estimate the source angular radius, $\theta_*= 0.81 \pm 0.07  \,\mu as$, where
the error includes uncertainty in the color conversion and the color/brightness-radius
relation.
On the other hand, ($V$-$I$, $I$)$_{\rm s,0}$ and optical color/brightness-radius 
relation by \cite{kervella2008}
yields $\theta_*= 0.83 \pm 0.05  \,\mu as$, which is consistent with above.
We adopt the mean of these estimates,
\begin{equation}
\theta_*= 0.82 \pm 0.07  \,\mu as. 
\label{eq:thetas}
\end{equation}
The angular and projected Einstein radii,
and lens-source relative proper motion $\mu$ are estimated, respectively, as:

\begin{eqnarray}
\theta_{\rm E}&=& {\theta_*\over\rho} =   529 \pm   84 \,\mu as,  \label{eq:thetaE}\\ 
\hat{r}_{\rm E}&=& \theta_E\times D_s = [4.2 \pm 0.7] \left( {D_s\over8{\rm kpc}} \right) \, {\rm AU}. \\ 
\mu&=&  {\theta_E \over t_{\rm E} }= 3.5 \pm   0.6 \, \rm mas\,yr^{-1}. 
\label{eq:mu}
\end{eqnarray}
As shown in the top right panel of
Figure~\ref{fig:likelihood}, the measured value of
$\mu=3.5$ ${\rm mas\,yr^{-1}}$ is typical for the bulge lenses but smaller than
the typical value for disk lenses, 5-10 ${\rm mas\,yr^{-1}}$, although it is
not inconsistent with a disk lens.

\section{Keck AO Observation}
\label{sec:keck}
$H$ and $K$ AO images of the event were taken by the Keck telescope at 
HJD=2454332.77689 and 2454332.77977, respectively.  The magnification at the
time of the Keck images are taken is $A_{\rm Keck}=2.490$.  
The magnified source position on the OGLE difference 
image is marked with the error of $\sim$2.5 pixels (25\,mas) in $K$-band.
From the Keck K-band image, the density of ambient stars with 3-$\sigma$ detection limit 
that correspond to $K\le$18.1 mag, is $\sim$0.3 arcsec$^{-2}$. We conservatively assume 
that the separation of two stars must be more than the measured FWHM of the PSF 
of 0.08 arcsec in order to be separately resolved.
Therefore the probability of blending with 
any random interloper, that is not related with this event, is only $\sim$0.6\%, 
implying this object is almost certainly the source, the lens and/or their companion.
The $H$ and $K$ magnitude were measured by PSF photometry and calibrated to 2MASS 
system using the $H$ and $K$ images taken by the IRSF telescope in South Africa,
following the method in \cite{Janczak10},

\begin{equation}
(H,K)_{\rm s,Keck}        =(  16.53 , 16.23) \pm (  0.03 ,  0.02), 
\label{eq:HKsKeck}
\end{equation}
and the magnitudes corrected for extinction given by $A_H/A_V=0.176$ and $A_K/A_V=0.105$,
which are estimated by using \cite{car89}'s law with the $R_V=2.64$ measured above, are
\begin{equation}
(H,K)_{\rm s,0,Keck}      =(  16.09 , 15.96 ) \pm (  0.04 ,  0.03). 
\label{eq:HKs0Keck}
\end{equation}

The $I-H$ and $I-K$ source colors are estimated from $(V-I)_{s,0}$ given by light 
curve fitting (Eq. \ref{eq:VIIs0}) by using the color-color relations of 
\cite{bessell88},
\begin{equation}
(I-H,I-K)_{\rm s,0} = ( 0.86_{-0.12}^{+0.11}, 0.92_{-0.13}^{+0.12} ). 
\label{eq:IHIKs0fit}
\end{equation}
Therefore $H$ and $K$ source magnitudes are given as,
\begin{equation}
(H,K)_{\rm s,0} = ( 17.23_{-0.19}^{+0.20}, 17.16\pm 0.20 ). 
\label{eq:HKs0fit}
\end{equation}
Then, the magnitude of the source when the Keck images were taken are
\begin{equation}
(H,K)_{\rm s,0} - 2.5({\rm log}[A_{\rm Keck}], {\rm log}[A_{\rm Keck}])= ( 16.23_{-0.19}^{+0.20}, 16.17\pm0.20 )  
\label{eq:HKs0fit+amp}
\end{equation}

By subtracting Eq. (\ref{eq:HKs0fit+amp}) from Eq. (\ref{eq:HKs0Keck}), we have the magnitude of 
the lens and/or companion of the lens or source, which serve as an upper mass limit of the lens
\begin{equation}
(H,K)_{l,\rm max,0}   =(  18.3_{-0.9}^{+\infty},   17.9_{-0.7}^{+\infty} ). 
\end{equation}
This $K$-band magnitude implies that the upper limit of the lens is an early G dwarf with mass of 
$M_{l,\rm max}= 1.0^{+0.2}_{-\infty} M_\sun$  
from \citep{schmidt-kaler82,bessell88}.
These $H$ and $K$-band upper limits are used for constraining lens star in Section \ref{sec:LensMass}.
If we could obtain a second epoch AO observation that gave us the baseline photometry, 
we would be able to constrain $(H,K)_{l,\rm max,0}$, much better.

For other (brighter) events, we have found that the
$H$-band source magnitude estimated by fitting the CTIO $H$-band light curve gives
a more precise value for the $H$ magnitude of the source. But, when we attempt
such an analysis for this event, we find significant indications of systematic errors.
This is not very surprising because this target does
not reach high magnification and is heavily blended with 
a nearby bright star. Also, because of the bright infrared sky brightness,
the CTIO $H$-band images do not go as deep as the optical images. Therefore we do not use
this CTIO $H$-band source magnitude in the following analysis.

\section{Lens System Masses and Distance}
\label{sec:LensMass}
The lens system mass, $M_l$, distance, $D_l$, and
lens-source relative velocity are directly constrained by only two
measured parameters, the
Einstein radius crossing time, $t_{\rm E}$, and the angular Einstein
radius, $\theta_E$. However we can further constrain them by 
a Bayesian analysis using a model of Galactic kinematics  
\citep{alc95,bea06,gou06,bennett08}.  The mass of the planet can be determined 
to the same precision as $M_l$ because the uncertainty in the mass ratio, $q$, 
is much smaller than the uncertainty in the Bayesian estimate of $M_l$.

\subsection{Planetary System Parameter for OGLE-2007-BLG-368Lb}
\label{sec:OGLE-2007-BLG-368Lb}

For this event, we observed finite source effects from which
we measured the angular Einstein radius $\theta_{\rm E}$ (Eq. \ref{eq:thetaE}), or 
equivalently the proper motion $\mu$ (Eq. \ref{eq:mu}), of the lens system.  
So we can break one link of the three-fold degeneracy by the relation,
\begin{equation}
\theta_E^2 = \kappa M \pi_{\rm rel}.
\label{eq:ML_finite}
\end{equation}

To produce the likelihood distributions shown in 
Figure \ref{fig:likelihood}, we compute the likelihood by combining this equation and
the measured values of $\theta_{\rm E}$ and $t_{\rm E}$ with the
Galactic model \citep{han03} assuming the distance to the GC is 8 kpc.
Here systematic errors in parameters estimated in Section \ref{sec:systematics} 
are taken into account.
This analysis yields that the primary is a K-dwarf with mass of 
$M_l = 0.64_{-0.26}^{+0.21}$ $M_\sun$ at 
$D_l = 5.9_{-1.4}^{+0.9}$ kpc and a planetary mass of
$M_p = 6.1_{-2.4}^{+2.0} \times 10^{-5}$ $M_\sun = 20_{-8}^{+7}$  $M_\oplus$ 
and projected separation of
$r_\perp   = 2.8_{-0.6}^{+0.5}$ AU.
The physical three dimensional separation $a = 3.3_{-0.8}^{+1.4}$ AU, can be estimated statistically
by putting a planetary orbit at random inclination and phase \citep{GouldLoeb1992}.
The lens-source relative proper motion $\mu   = 3.3_{-0.3}^{+0.4}$ mas\,yr$^{-1}$,
which is consistent with the value given by Eq. (\ref{eq:mu}), favors that the lens 
is in the bulge rather than the disk in which typically $\mu=$ 5-10 mas\,yr$^{-1}$.

\subsection{Comparison to Other Known Exoplanets}
\label{sec:comparison}

Figures \ref{fig:MvsSemiMajor} and \ref{fig:MvsSnowLine} compare the
masses and semi-major axes of the planets found by microlensing to those
found by other methods. Figure \ref{fig:MvsSnowLine} takes into account
the different masses of the primary stars and uses the ratio of the
semi-major axis to the position of the snow-line as the x-axis parameter
in order to display the data in a way more relevant to planet formation
theories.

The positions of all the microlensing planets on these plots are determined
by a Bayesian analysis similar to the one we present for OGLE-2007-BLG-368Lb.
However, there is a crucial distinction. The events plotted with red filled circle
and error bars have masses determined either by microlensing parallax measurements
or by direct identification of the lens star in HST images, so these
can be considered to be actual measurements. The other microlensing planets,
plotted with red open circle and error bars, are like OGLE-2007-BLG-368Lb, in that the
light curve measurements do not directly determine the lens system mass.
For these events, the derived parameters have a significant dependence on
the assumed prior, and we must be careful not to over-interpret the results.
For example, we cannot use the results of these Bayesian analyses to study
the probability that a star will host a planet as a function of its mass, because
these estimates of the host star mass depend upon our prior assumptions
about this planet hosting probability. Instead, such questions must be studied
with a new Bayesian analysis using only directly measured quantities as
constraints.

There are planetary microlensing events that warrant some additional discussion.
The Bayesian analyses for these events yield double-peaked likelihood 
functions. This gives results that are extremely sensitive to the prior assumptions,
so one should not directly use the Bayesian results in these cases.
Event MOA-2007-BLG-400 has a severe $d \leftrightarrow 1/d$
model degeneracy, which yields a factor of 10 uncertainty in the projected
star-planet separation. We extend the error bars from the 1-$\sigma$
lower limit on the semi-major axis from the $d < 1$ solution to the
1-$\sigma$ upper limit from the $d > 1$ solution.

The other ambiguous planet is MOA-2008-BLG-310Lb \citep{Janczak10}.
This event is unusual because the kinematics favors a low-mass primary of
$\sim 0.1 \msun$, while the excess flux seen in VLT/NACO images of the source 
star suggests a much more massive planetary host star
with $M \sim 0.7\msun$. But this excess flux 
could be due to a companion to the lens, source, or the
chance superposition of an unrelated star. So, the Bayesian analysis
yields two peaks for the lens star (and planet) masses, but the
relative weighting of these two peaks is quite sensitive to the
assumed prior. So, as with MOA-2007-BLG-400, we use the 1-$\sigma$
upper and lower limits on the high-mass and low-mass solutions for
our error bars for this event. For the central point, we use the
geometric mean of the peaks of the high-mass and low-mass solutions.

\section{Constraints on the Planetary Mass Function}
\label{sec:mass_func}

In Figures~\ref{fig:MvsSemiMajor} and \ref{fig:MvsSnowLine}, it appears that the
distribution of planets found by microlensing is roughly independent
of mass above $1\mearth$, with perhaps a peak at $M \sim 10\mearth$.
However, the probability that a planet can be detected by microlensing
depends on its mass, and these figures have not been corrected
for the planet detection efficiency \citep{albrow00,Gaudi2000,rhie00}. A full calculation of the planet detection
efficiency (Cassan et al., in preparation) including detail assessments of various 
potential systematics is beyond the scope of this paper,
but we can obtain interesting constraints on the planetary mass 
function using a simple model for the relative planet detection efficiency.

For events with signals due to the planetary caustic
\citep{GouldLoeb1992}, there are some simple theoretical
arguments regarding the dependence of the planet detection efficiency
on the mass ratio, $q$. If we ignore finite source effects, which are usually
unimportant for planets with masses $\simgt 10\mearth$
\citep{bennett96}, then the planetary caustic shape is nearly independent
of $q$, and its area scales as $q$. We can define a planet detection
region as the area of the lens magnification pattern that differs from the
single lens light curve \citep{pac86} by more than some threshold (either
relative or absolute). With such a definition, the area of the planet detection
region will scale as the size of the planetary caustic, as $q$.
Then the probability that a given source trajectory will cross the planet
detection region scales as the linear dimension of this region, which goes
as $q^{1/2}$. So, in the limit of very good light curve coverage, the planet
detection efficiency for planetary caustic events should scale as $q^{1/2}$.
However, the duration of the planetary deviation also scales as $q^{1/2}$,
and with sparse light curve coverage or large photometric error bars, 
the detection efficiency can scale as steeply as $q$. In general, we expect
that situation to be in between these extremes for planetary caustic 
events, so that we should expect that the planet detection efficiency
should have some scaling intermediate between $q^{1/2}$ and 
$q$. We have calculated the detection efficiency for OGLE-2007-BLG-368
using the method of \citet{rhie00}. Of course, it would be inconsistent
to use the follow-up data that was taken because the planetary signal
was recognized in such a calculation, so we have only included the
regularly scheduled survey data in this calculation. 
The dependence of the detection efficiency on the detection threshold 
in different alert systems is negligible compared to the dependence on 
the lightcurve coverage of the dataset. This calculation gives a 
detection efficiency for OGLE-2007-BLG-368 that scales as $\sim q^{0.8}$
at the range of $q$ appropriate for Neptune-mass to Jupiter-mass planets.
This same scaling also holds true for the two other microlensing planets
discovered through planetary caustic deviations, OGLE-2003-BLG-235,
and OGLE-2005-BLG-390 \citep{kubas08}. For all the calculations in
this section, we assume that the distribution of planets is uniform in
$\log(d)$ for all separations, $d$, and we sum over all separations.

The situation is somewhat different for high magnification microlensing
events, which are detected through perturbations of the central caustic.
Since the linear size of the central caustic scales as $q$ \citep{dominik99}, 
one might expect that the detection efficiency would scale more steeply
than $q$ for the same reasons that the planetary caustic planet detection
efficiency scales more steeply than $q^{1/2}$, but this is not the case.
The reason for this is that for events of sufficiently high magnification,
$A_{\rm max} > 50$ or so, the planet detection efficiency for Jupiter-mass
planets saturates at 1 for a wide range of separations. This is, in fact,
the main reason why the observing groups focus on high magnification events
\citep{griest98}. The planet detection efficiency has been calculated for
a number of high magnification events
\citep{albrow01,gau02,rat02,abe04,yoo04b,dong-ogle343,bennett08,
nagaya09,yee09,batista09}, and these events reveal detection efficiency
scalings that range from $q^{0.7}$, for MOA-2006-BLG-130 and MOA-2007-BLG-192,
to $q^{0.3}$ for OGLE-2008-BLG-279, which is the event with the 
highest planet detection sensitivity \citep{yee09}. Generally, the scaling is
shallowest for the events with the highest sensitivity to planets and steeper for
events with lower sensitivity due to lower peak magnification, less complete
light curve sampling, or less precise photometry. For the collection of
high magnification events observed, we estimate that the mean detection
efficiency scales as $q^{0.5\pm 0.1}$, and for all microlensing events,
we estimate that the detection efficiency scales as $q^{0.6\pm 0.1}$.

We can now use the detection efficiency estimate to infer some properties 
of the distribution of planets in our Galaxy. In analogy to the stellar 
mass function, we define the planetary mass ratio function, $dN_{\rm pl}/d \log q$,
such that the number of planets per star in a logarithmic mass ratio interval is
given by $dN_{\rm pl}/d \log q$.
We assume to have a power-law form for the planetary
mass ratio function, 
\begin{equation}
{dN_{\rm pl}\over d \log q} = N_0\, q^n \,\Theta(q-q_0) \Theta(q_1-q)  \ ,
\label{eq:PMF}
\end{equation}
where $q_0$ and $q_1$ are the
lower and upper limits on the planetary mass ratio. ($q_0$ could alternatively
be considered to be a low-mass-ratio cutoff on the planetary detection efficiency.)
Thus, $n = 0$ would imply that there are an equal number of planets in
every logarithmic mass interval, and $n = -1$ would imply that total
mass of planets in every logarithmic mass interval is the same.

\begin{figure}
\epsscale{0.5}
\includegraphics[angle=0,scale=0.75,keepaspectratio]{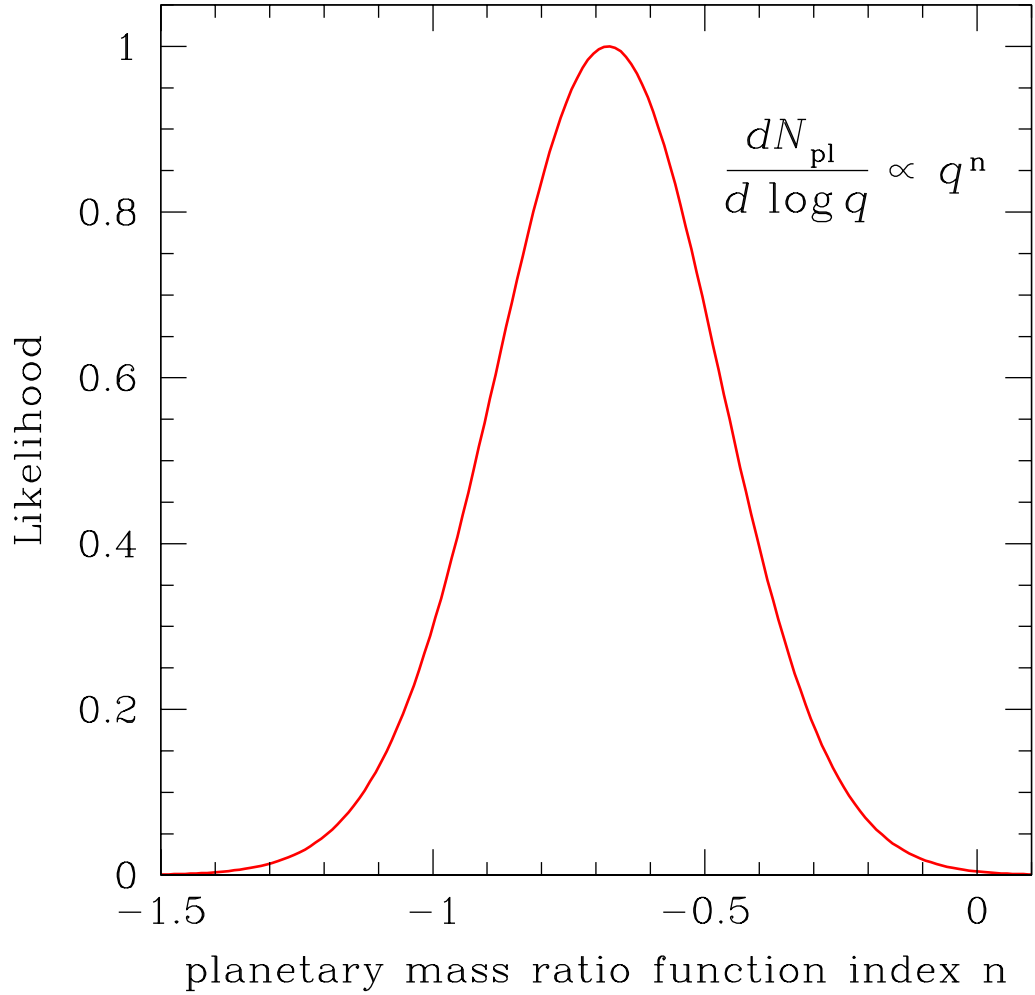}
\caption{
The probability distribution of the power law  index, $n$, of the
planetary mass ratio function, $\Psi(q)$, based upon the mass ratios of 
the ten exoplanets detected by microlensing and our estimate of the 
planetary detection efficiency. This calculation yields
$n = -0.68 \pm 0.20$, with a 95\% confidence level upper limit of
$n < -0.35$.
\label{fig:eff_pow}}
\end{figure}

We can estimate the parameters, $N_0$ and $n$ that describe the
planetary mass ratio function using a likelihood analysis.
The expression for the likelihood function for the
planetary mass ratio function parameters is just the Poisson probability 
of finding the observed number of events, $N_{\rm obs}$, times the
product of the probability of finding events which each of the observed
mass ratios, $q_i$. This can be written as
\begin{equation}
{\cal L}(N_0,n)  = e^{-N_{\rm exp}} \prod_i^{N_{\rm obs}}{dN_{\rm pl}\over d \log q}{\cal E}(q_i)  \ ,
\end{equation}
where ${\cal E}(q) \propto q^{0.6\pm 0.1}$ is the planet detection efficiency
and $N_{\rm exp}$ is the number of events expected for the given
$N_0$ and $n$ values \citep{macho-lmc1,macho-lmc2}. However, 
since we have only calculated relative and not absolute efficiencies, we cannot
calculate $N_{\rm exp}$ and we cannot hope to constrain $N_0$.
Therefore, we adjust $\Phi_0$ so that $N_{\rm exp} = N_{\rm obs}$, and
evaluate the likelihood function for only the power-law index, $n$, of the
planetary mass ratio function. The resulting likelihood function based on the
ten planets discovered by microlensing is shown in 
Figure~\ref{fig:eff_pow}, and the resulting planetary mass ratio function
index is $n = -0.68 \pm 0.20$, with a 95\% confidence level upper 
limit of $n < -0.35$. 
The core of this distribution is similar to a
Gaussian, but the distribution is skewed, with a higher probability
of a $>2$-$\sigma$ deviation at small $n$ than at large $n$
This error bar includes the $\pm 0.1$ uncertainty in the 
detection efficiency power law index (${\cal E}(q) \propto q^{0.6\pm 0.1}$).
This result does have some dependence on our choice of the lower and
upper cutoffs of $q_0 = 3\times 10^{-5}$ and $q_1 = 0.015$, but the
variation due to the choice of these cutoffs is much smaller than the resulting
uncertainty in $n$.

This result for the power law index indicates that we should expect
$7{+6\atop -3}$ times as many cold Neptunes ($q\sim 5\times 10^{-5}$) 
as Jupiters ($q\sim 10^{-3}$), with a 
95\% confidence level lower limit of 2.8 times as many cold Neptunes
as Jupiters. This is in line with the basic predictions of the core
accretion model \citep{ida04,lau04}, as these models predict 
a large population of Neptune-like, ``failed Jupiter" cores to form
beyond the snow line, particularly for stars of less than a solar mass,
which make up most of the sample probed by microlensing. However,
it still may be possible to explain this result in the context of the 
gravitational instability theory \citep{boss06}.

This power law index of $n = -0.68 \pm 0.20$ is steeper than
(but consistent with) the
index of $n = -0.31\pm 0.20$ found by \citet{cumming08} for
more massive planets orbiting mostly solar-type stars. This is 
also steeper than the mass function prediction of 
\citet{mord09} for solar-type stars, although this theoretical
mass function is not a power law. Radial velocity surveys also
find that hot Neptunes, with periods less than 50 days
are quite common around G and K dwarfs \citep{may09}.

\section{Discussion and Conclusions}
\label{sec:conclusion}

We have presented the analysis of the OGLE-2007-BLG-368 planetary
microlensing event, which indicates that the planet OGLE-2007-BLG-368Lb
is a Neptune-mass planet.  We also find evidence for low level systematic
errors in the light curve, which however do not affect this conclusion.
We estimate the systematic errors by taking the differences between the 
various models, i.e., the standard and xallarap with and without Kepler 
constraint.

\begin{figure}
\epsscale{0.5}
\includegraphics[angle=0,scale=0.75,keepaspectratio]{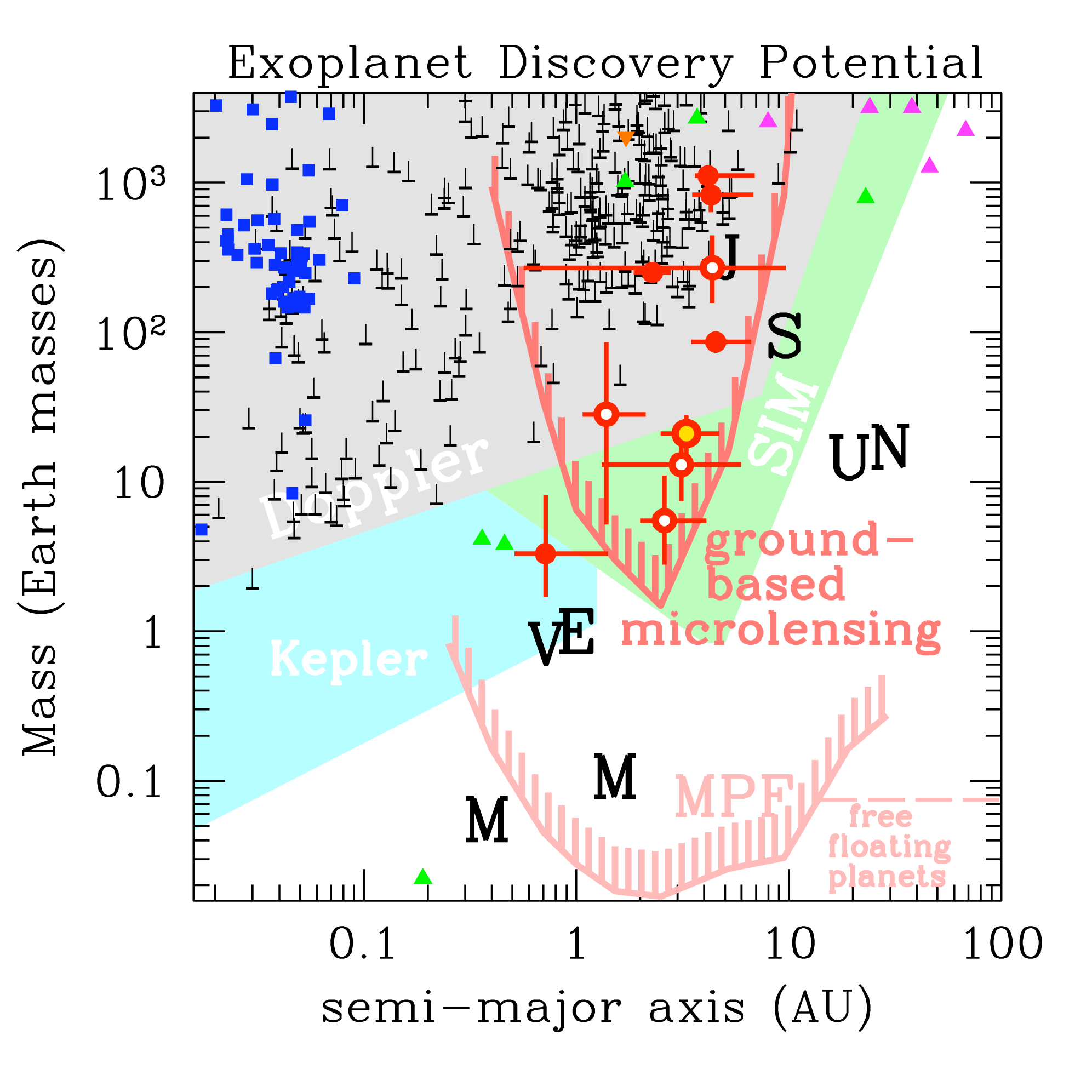}
\caption{
Known exoplanets as a function of mass vs.\ semi-major axis, along with the 
predicted sensitivity curves for various methods.
The red filled and open circles with error bars indicate the microlensing planets
with mass measurements and mass estimated by Bayesian analysis, respectively
(see section \ref{sec:comparison}). OGLE-2007-BLG-368Lb is indicated
by the gold-filled open circle.
The blue dots represent the planets first detected via transit. 
The black bars with upward-pointing error bars (indicating $1\,\sigma$ $\sin i$ uncertainty) 
are the radial velocity planet detections.
The green and magenta triangles indicate the planets found by timing 
(including the pulsar planets) and by direct detection, respectively.
The yellow, cyan, and light green shaded regions indicate the expected 
sensitivity limits of the radial velocity, {\it Kepler} and {\it SIM} space missions.
The red and pink  curves indicate the predicted lower sensitivity limits for a
ground-based and space-based \citep{bennett02} microlensing planet search 
program, respectively. The solar system's planets are indicated with black letters.
\label{fig:MvsSemiMajor}}
\end{figure}

\begin{figure}
\epsscale{0.5}
\includegraphics[angle=0,scale=0.75,keepaspectratio]{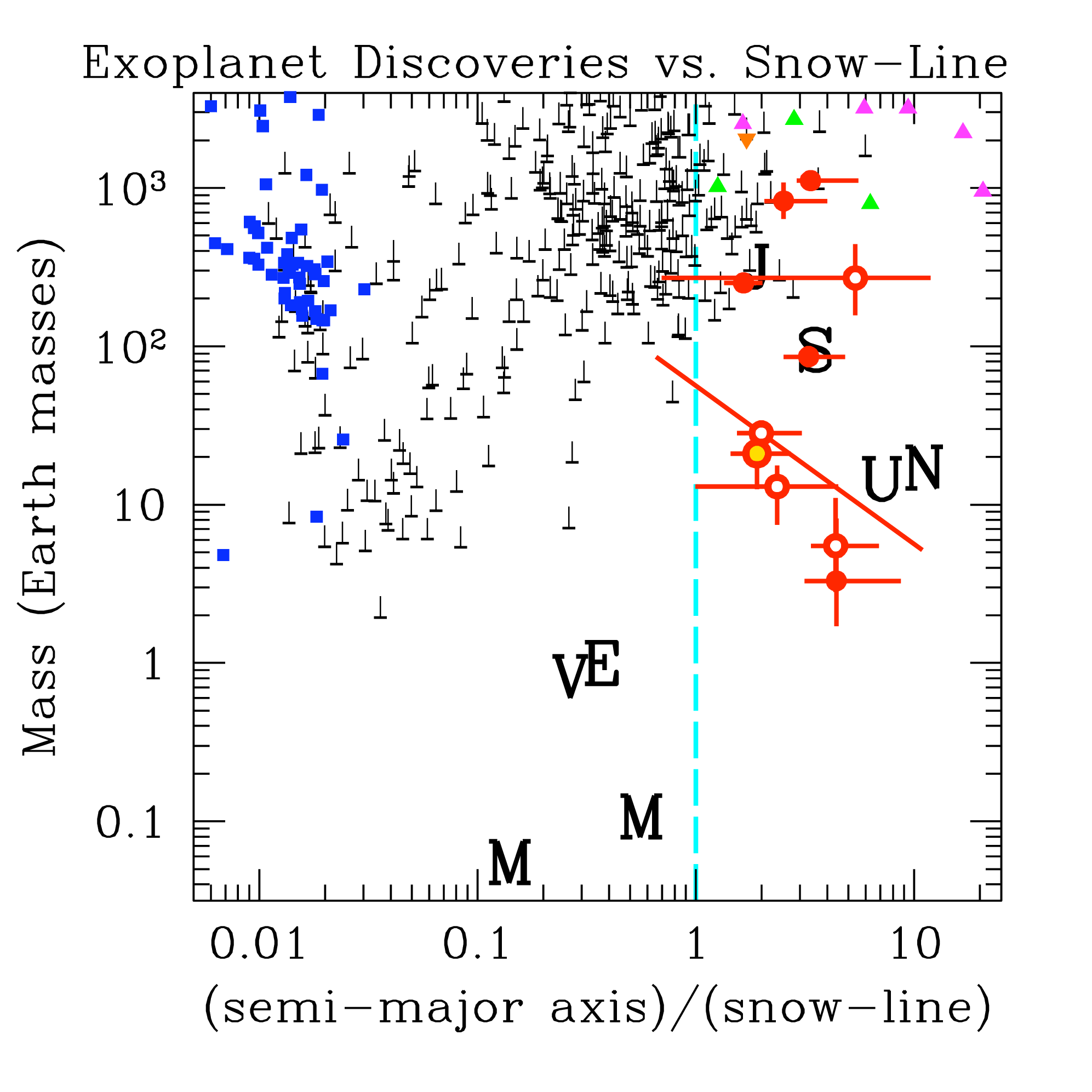}
\caption{
Known exoplanets as a function of mass vs.\ semi-major axis divided by the
snow line, which is taken to be at $a_{\rm snow}=2.7$ AU $M/M_\sun$. 
As in Figure \ref{fig:MvsSemiMajor}, microlensing planets are indicated by red 
filled and open circles with error bars (see section \ref{sec:comparison}). 
OGLE-2007-BLG-368Lb is indicated
by the gold-filled open circle.
Blue dots represent the planets first 
detected by transits.  The black bars with upward-pointing error bars are the 
planets detected via the radial velocity.
The green and magenta triangles indicate the planets found by timing 
(including the pulsar planets) and by direct detection, respectively.
\label{fig:MvsSnowLine}}
\end{figure}

By using a Bayesian analysis, we found the planet has a mass of 
$M_p = 20_{-8}^{+7}$ $M_\oplus$ and a projected separation of 
$r_\perp   = 2.8_{-0.6}^{+0.5}$ AU around a K-dwarf with mass of 
$M_l = 0.64_{-0.26}^{+0.21}$ $M_\sun$ at 
$D_l = 5.9_{-1.4}^{+0.9}$ kpc.
This is the 4th Neptune-mass planet detected by microlensing.
In Figure \ref{fig:MvsSemiMajor}, we plot these planets as a function of 
mass vs.\ semi-major axis along with all known exoplanets. 
Figure \ref{fig:MvsSnowLine} is the same as Figure \ref{fig:MvsSemiMajor}
but the semi-major axis is divided by the snow line, which is taken to be at 
$a_{\rm snow}=2.7$ AU $M/M_\sun$.
As for the microlensing planets in this figure, we are starting to see a 
broad concentration 
of $\sim$ 10 $M_\oplus$ planets beyond the snow line. 
This is as expected from the core accretion theory. This theory predicts 
that the most massive solid planetary cores should form beyond the snow line, which then 
accrete nebular gas and become the gas giants around the solar-type star. 
On the other hand, they become Earth-mass to Neptune-mass icy rocky 
planets around M-dwarfs.  
Comparing four Neptune-mass, five Jovian planets and one between Neptune and Saturn
found by microlensing, it confirms that cold Neptunes are relatively common 
around low mass primary stars analyzed by \cite{gou06}.

We have presented an analysis of the exoplanet mass ratio function. Assuming
that the number of planets scales as a power law in the mass ratio, $q$, we
define the mass ratio function as 
${dN_{\rm pl}/ d \log q} \propto q^n$ power law mass 
function, over the mass range of a few Jupiter masses down to a few Earth-masses,
we find a power law index of $n = -0.68\pm 0.20$, which indicates that
Neptune-mass planets are substantially more common than Jupiter-mass planets.

The planetary signature of this event was detected in real-time in the data points
from survey telescopes MOA and OGLE. Then the signature was greatly clarified by intensive 
follow-up observations prompted by the alert.
This is a planetary caustic crossing event, the second of its kind 
after OGLE 2005-BLG-390 \citep{bea06} among all planetary microlensing events. 
OGLE 2003-BLG-235 \citep{bon04} crossed the planetary part of a resonant caustic.
Although the time of the planetary deviation in these events can not be predicted 
for planetary-caustic events, the potential
event rate is higher than central-caustic event in which the time of the
planetary deviation is known \citep{han01}. This discovery shows that the high cadence survey 
observations that MOA is conducting, have a great potential to increase the event 
rate of the planetary microlensing.
In 2010, OGLE will upgrade its camera to 1.4 deg$^2$ FOV (OGLE-IV), which will
enable OGLE to follow a similar strategy of high cadence monitoring for planetary signals.

Multi-continent high-cadence observing will commence in 2010
with the start of the OGLE-IV project, and in future years will
expand further when the Korean Microlensing Telescope Network (KMTNet)
is commissioned. These improvements can be expected to dramatically 
increase the number of microlensing planets, and in particular those 
like OGLE-2007-BLG-368Lb, that are discovered via planetary-caustic 
perturbations.

\vspace{5cm}

\acknowledgments 
This work is supported by the grant JSPS18253002 and JSPS20340052 (MOA).
TS was supported by MEXT Japan, Grant-in-Aid for Young Scientists (B), 18749004 
and Grant-in-Aid for Scientific Research on Priority Areas, "Development of 
Extra-solar Planetary Science", 19015005.
D.P.B.\ was supported by grants AST-0708890 from the NSF and NNX07AL71G from NASA.
The OGLE project is partially supported by the Polish MNiSW grant
N20303032/4275 to AU.
Work by A.G. was supported by NSF grant AST-0757888.
Work by B.S.G., A.G. and R.P. supported by NASA grant NNG04GL51G
Dr Dave Warren provides financial support for Mt Canopus Observatory.
CH was supported by Creative Research Initiative Program (2009-0081561)
of National Research Foundation of Korea (CH). B-GP and C-UL were
supported by the grant of Korea Astronomy and Space Science Institute.

\clearpage

\begin{deluxetable}{lrrrrrrrrrrrrrr}
\rotate
\tablecaption{Model parameters.  \label{tbl:param}}
\tablewidth{0pt}
\tablehead{
\colhead{model}         & \colhead{$t_0$}          & \colhead{$t_{\rm E}$}    & \colhead{$u_{0}$} &
\colhead{$q$}           & \colhead{$d$}            & \colhead{$\alpha$}       & \colhead{$\rho$}        &
\colhead{$\pi_{\rm E}$} & \colhead{$\phi_{\rm E}$} & \colhead{$\xi_{\rm E}$}  &
\colhead{$\phi_{\xi}$}  & \colhead{$P_\xi$}        & \colhead{$\epsilon$}     & \colhead{$\chi^2$}           \\
\colhead{}              & \colhead{HJD$'$}         & \colhead{days}           & \colhead{}              &
\colhead{$10^{-4}$}     & \colhead{}               & \colhead{rad}            & \colhead{$10^{-3}$}     &
\colhead{}              & \colhead{rad}            & \colhead{}               &
\colhead{rad}           & \colhead{days}           & \colhead{}               & \colhead{}
}
\startdata
standard     & 4310.92 & 53.2 & 0.0825 & 1.27 & 0.9227 & 0.452 & 1.88 &  --- &  --- &  --- &  --- &   --- &   --- & 3306.3\\
$\sigma$     &    0.01 &  0.4 & 0.0008 & 0.02 & 0.0006 & 0.002 & 0.03 &  --- &  --- &  --- &  --- &   --- &   --- & --- \\
parallax     & 4311.07 & 59.9 & 0.0765 & 0.77 & 0.9286 & 0.534 & 1.39 & 1.78 & 5.66 &  --- &  --- &   --- &   --- & 3008.4\\
$\sigma$     &    0.02 &  1.0 & 0.0011 & 0.04 & 0.0009 & 0.008 & 0.04 & 0.14 & 0.04 &  --- &  --- &   --- &   --- & --- \\
xallarap$^*$ & 4311.10 & 54.1 & 0.0790 & 0.89 & 0.9257 & 0.462 & 1.52 &  --- &  --- & 1.73 & 6.08 & 215.9 &   --- & 2934.9\\
$\sigma$ &        0.01 &  0.5 & 0.0008 & 0.02 & 0.0007 & 0.002 & 0.03 &  --- &  --- &  --- &  --- &   --- &   --- & --- \\
xallarap$^{*K}$&4311.12& 53.2 & 0.0796 & 0.99 & 0.9252 & 0.438 & 1.61 &  --- &  --- & 0.21 & 6.18 & 102.4 &   --- & 2975.7\\
$\sigma$ &        0.01 &  0.6 & 0.0010 & 0.02 & 0.0008 & 0.002 & 0.03 &  --- &  --- &  --- &  --- &   --- &   --- & --- \\
xallarap     & 4311.08 & 57.7 & 0.0781 & 0.85 & 0.9266 & 0.516 & 1.46 &  --- &  --- & 0.35 & 6.20 & 103.0 & 0.48  & 2919.0\\
$\sigma$     &    0.01 &  0.6 & 0.0009 & 0.02 & 0.0007 & 0.002 & 0.03 &  --- &  --- &  --- &  --- &   --- &   --- & --- \\
xallarap$^K$ & 4311.12 & 55.4 & 0.0793 & 0.95 & 0.9255 & 0.478 & 1.55 &  --- &  --- & 0.16 & 4.73 & 106.3 & 0.77  & 2936.9\\
$\sigma$     &    0.01 &  0.5 & 0.0007 & 0.02 & 0.0006 & 0.002 & 0.03 &  --- &  --- &  --- &  --- &   --- &   --- & --- \\
\hline
$\sigma_{\rm systematic}$& 0.01 &  2.3 & 0.0022 & 0.21 & 0.0019 & 0.039 & 0.21 &  --- &  --- &  --- &  --- &   --- &   --- & --- \\
\enddata
\tablecomments{
Standard model includes neither parallax nor xallarap effects.  HJD$'$ = HJD-2450000.
Models with superscripts ``*'' and ``$K$'' indicate fixed $\epsilon=0$
and Kepler constraint $M_s = M_c = 1\,M_\odot$, respectively.
The lines with "$\sigma$" list the 1-$\sigma$ error of parameters given by MCMC, for which
the xallarap parameters are fixed at the best values for the xallarap models because
xallarap parameters are strongly degenerate and it is hard to satisfy the convergence criteria.
$\sigma_{\rm systematic}$ indicates the systematic errors (see Section \ref{sec:systematics}).
}
\end{deluxetable}

\end{document}